\pdfoutput=1

\documentclass[aps,twocolumn,groupedaddress,floatfix,preprintnumbers,nofootinbib,eqsecnum]{revtex4}

\usepackage{amsmath, float, graphicx,placeins,amssymb,multirow,pgfplots,pgfplotstable}
\usepackage{tikz}
\usepackage{slashed}
\usetikzlibrary{shapes.geometric,arrows}
\graphicspath{{./Plots/}}

\begin{document}
\title{
Off-Diagonal Dark-Matter Phenomenology: \\ 
Exploring Enhanced Complementarity Relations \\
in Non-Minimal Dark Sectors 
}

\author{Keith R.\ Dienes$^{1,2}$\footnote{E-mail address:  {\tt dienes@email.arizona.edu}},
      Jason Kumar$^{3}$\footnote{E-mail address:  {\tt jkumar@hawaii.edu}},
      Brooks Thomas$^{4}$\footnote{E-mail address:  {\tt thomasbd@lafayette.edu}},
      David Yaylali$^{1}$\footnote{E-mail address:  {\tt yaylali@email.arizona.edu}}}
\affiliation{
     $^1$Department of Physics, University of Arizona, Tucson, AZ  85721  USA\\
     $^2$Department of Physics, University of Maryland, College Park, MD  20742  USA\\
     $^3$Department of Physics \& Astronomy, University of Hawaii, Honolulu, HI 96822  USA\\
     $^4$Department of Physics, Lafayette College, Easton, PA 18042  USA}

\begin{abstract}
In most multi-component dark-matter scenarios, two classes of processes generically 
contribute to event rates at experiments capable of probing 
the nature of the dark sector.  The first class consists of ``diagonal'' processes 
involving only a single species of dark-matter particle --- processes 
analogous to those which arise in single-component dark-matter scenarios.    
By contrast, the second class consists of ``off-diagonal'' processes involving 
dark-matter particles of different species.  Such processes include inelastic 
scattering at direct-detection experiments, asymmetric production at colliders,
dark-matter co-annihilation, and certain kinds of dark-matter decay. 
In typical multi-component scenarios, the contributions from diagonal processes 
dominate over those from off-diagonal processes.
Unfortunately, this tends to mask those features which are most sensitive to the multi-component
nature of the dark sector.  
In this paper, by contrast, we point out that there exist natural, multi-component
dark-sector scenarios in which the off-diagonal contributions
actually dominate over the diagonal.
This then gives rise to a new, enhanced
picture of dark-matter complementarity.   In this paper, we introduce a scenario in which this 
situation arises and examine the enhanced picture of dark-matter 
complementarity which emerges.
\end{abstract}

\maketitle


\newcommand{\PRE}[1]{{#1}} 
\newcommand{\ul}{\underline}
\newcommand{\del}{\partial}
\newcommand{\nbox}{{\,\lower0.9pt\vbox{\hrule \hbox{\vrule height 0.2 cm
\hskip 0.2 cm \vrule height 0.2 cm}\hrule}\,}}

\newcommand{\postscript}[2]{\setlength{\epsfxsize}{#2\hsize}
   \centerline{\epsfbox{#1}}}
\newcommand{\gweak}{g_{\text{weak}}}
\newcommand{\mweak}{m_{\text{weak}}}
\newcommand{\mplanck}{M_{\text{Pl}}}
\newcommand{\mstar}{M_{*}}
\newcommand{\sigmaan}{\sigma_{\text{an}}}
\newcommand{\sigmatot}{\sigma_{\text{tot}}}
\newcommand{\sigmaSI}{\sigma_{\rm SI}}
\newcommand{\sigmaSD}{\sigma_{\rm SD}}
\newcommand{\OmegaM}{\Omega_{\text{M}}}
\newcommand{\OmegaDM}{\Omega_{\text{DM}}}
\newcommand{\ipb}{\text{pb}^{-1}}
\newcommand{\ifb}{\text{fb}^{-1}}
\newcommand{\iab}{\text{ab}^{-1}}
\newcommand{\ev}{\text{eV}}
\newcommand{\kev}{\text{keV}}
\newcommand{\mev}{\text{MeV}}
\newcommand{\gev}{\text{GeV}}
\newcommand{\tev}{\text{TeV}}
\newcommand{\pb}{\text{pb}}
\newcommand{\mb}{\text{mb}}
\newcommand{\cm}{\text{cm}}
\newcommand{\m}{\text{m}}
\newcommand{\km}{\text{km}}
\newcommand{\kg}{\text{kg}}
\newcommand{\g}{\text{g}}
\newcommand{\s}{\text{s}}
\newcommand{\yr}{\text{yr}}
\newcommand{\days}{\text{days}}
\newcommand{\Mpc}{\text{Mpc}}
\newcommand{\etal}{{\em et al.}}
\newcommand{\eg}{{\em e.g.}}
\newcommand{\ie}{{\em i.e.}}
\newcommand{\ibid}{{\em ibid.}}
\newcommand{\Eqref}[1]{Equation~(\ref{#1})}
\newcommand{\secref}[1]{Sec.~\ref{sec:#1}}
\newcommand{\secsref}[2]{Secs.~\ref{sec:#1} and \ref{sec:#2}}
\newcommand{\Secref}[1]{Section~\ref{sec:#1}}
\newcommand{\appref}[1]{App.~\ref{sec:#1}}
\newcommand{\figref}[1]{Fig.~\ref{fig:#1}}
\newcommand{\figsref}[2]{Figs.~\ref{fig:#1} and \ref{fig:#2}}
\newcommand{\Figref}[1]{Figure~\ref{fig:#1}}
\newcommand{\tableref}[1]{Table~\ref{table:#1}}
\newcommand{\tablesref}[2]{Tables~\ref{table:#1} and \ref{table:#2}}
\newcommand{\met}{\not{\! \! E_T}}
\newcommand{\Dsle}[1]{\slash\hskip -0.28 cm #1}
\newcommand{\Dslp}[1]{\slash\hskip -0.23 cm #1}
\newcommand{\mpt}{{\Dslp p_T}}
\newcommand{\Dsl}[1]{\slash\hskip -0.20 cm #1}

\newcommand{\mB}{m_{B^1}}
\newcommand{\mq}{m_{q^1}}
\newcommand{\mf}{m_{f^1}}
\newcommand{\mKK}{m_{KK}}
\newcommand{\WIMP}{\text{WIMP}}
\newcommand{\SWIMP}{\text{SWIMP}}
\newcommand{\NLSP}{\text{NLSP}}
\newcommand{\LSP}{\text{LSP}}
\newcommand{\mWIMP}{m_{\WIMP}}
\newcommand{\mSWIMP}{m_{\SWIMP}}
\newcommand{\mNLSP}{m_{\NLSP}}
\newcommand{\mchi}{m_{\chi}}
\newcommand{\mgravitino}{m_{\gravitino}}
\newcommand{\mmed}{M_{\text{med}}}
\newcommand{\gravitino}{\tilde{G}}
\newcommand{\Bino}{\tilde{B}}
\newcommand{\photino}{\tilde{\gamma}}
\newcommand{\stau}{\tilde{\tau}}
\newcommand{\slepton}{\tilde{l}}
\newcommand{\snu}{\tilde{\nu}}
\newcommand{\squark}{\tilde{q}}
\newcommand{\mgaugino}{M_{1/2}}
\newcommand{\epsEM}{\varepsilon_{\text{EM}}}
\newcommand{\mmess}{M_{\text{mess}}}
\newcommand{\lmess}{\Lambda}
\newcommand{\nmess}{N_{\text{m}}}
\newcommand{\signmu}{\text{sign}(\mu)}
\newcommand{\Omegachi}{\Omega_{\chi}}
\newcommand{\lambdafs}{\lambda_{\text{FS}}}
\newcommand{\be}{\begin{equation}}
\newcommand{\ee}{\end{equation}}
\newcommand{\bea}{\begin{eqnarray}}
\newcommand{\eea}{\end{eqnarray}}
\newcommand{\baln}{\begin{align}}
\newcommand{\ealn}{\end{align}}
\newcommand{\lsim}{\lower.7ex\hbox{$\;\stackrel{\textstyle<}{\sim}\;$}}
\newcommand{\gsim}{\lower.7ex\hbox{$\;\stackrel{\textstyle>}{\sim}\;$}}
\newcommand{\Dslash}{\!\not{\!\! D}}

\def\beq{\begin{equation}}
\def\eeq{\end{equation}}
\def\beqn{\begin{eqnarray}}
\def\eeqn{\end{eqnarray}}

\newcommand{\ssection}[1]{{\em #1.\ }}
\newcommand{\rem}[1]{\textbf{#1}}

\newcommand{\vmin}{v_{\text{min}}}
\newcommand{\vmax}{v_{\text{max}}}

\def\ie{{\it i.e.}\/}
\def\eg{{\it e.g.}\/}
\def\etc{{\it etc}.\/}

\def\sqtsn{\sqrt{s_n}}
\def\sqts0{\sqrt{s_0}}
\def\Dsqts{\Delta(\sqrt{s})}
\def\Omegatot{\Omega_{\mathrm{tot}}}
\def\tnow{t_{\mathrm{now}}}
\def\trec{t_{\mathrm{rec}}}
\def\rhotot{\rho_{\mathrm{tot}}}


\section{Introduction\label{sec:Introduction}}


Overwhelming evidence suggests that particle dark matter contributes a substantial 
fraction of the present-day energy density in the universe~\cite{DMReviews}.  All of 
this evidence ultimately relies upon gravitational interactions between the dark matter and 
visible matter.  However, it is possible that the fields 
constituting the dark matter may interact with the fields of the Standard Model (SM) 
in other ways as well.  This possibility has inspired a number of experimental 
strategies for observing dark matter, including direct detection, indirect 
detection, and collider searches.  

In typical theoretical dark-matter models, the 
underlying amplitudes which provide the leading contributions to dark-matter 
production, scattering, and annihilation rates are related by crossing symmetries.  
For this reason, different experimental probes of the dark sector can be seen as 
complementary in two ways.  First, these different probes explore different 
regions of the parameter space of a given dark-matter model.  Second, the comparison of 
results from multiple such probes can assist us in distinguishing between different 
models.  In this way, the complementarities between different experimental probes of 
the dark sector can provide a useful tool for exploring and constraining the parameter 
space of a particular dark-matter model (for a review, see, 
\eg, Ref.~\cite{ComplementarityWhitePaper}).  
     
In multi-component dark-matter scenarios, however, the experimental complementarity picture 
can be even richer and more subtle than it is in traditional, single-component scenarios~\cite{DecayComplementarity}.  
One reason is that the multi-component nature of the dark sector allows for two distinct
classes of processes which can generically contribute  
to the relevant experimental event rates.
The first class consists of ``diagonal'' processes 
involving only a single species of dark-matter particle.  These processes, which
are analogous to those which arise in single-component dark-matter scenarios, include
the elastic scattering of dark matter with visible matter, dark-matter annihilation,
and symmetric pair-production processes at colliders.  By contrast, the second class 
consists of ``off-diagonal'' processes involving two dark-matter particles of different 
species.  These processes include the inelastic scattering of dark matter with visible 
matter, dark-matter co-annihilation, and asymmetric pair-production processes at
colliders.  Moreover, this latter class also includes a wholly different 
type of process that has no analogue in single-component dark-matter models:
the decay of a heavier dark-matter species to a final state comprising a lighter
dark-matter species and some number of SM particles~\cite{DecayComplementarity}.  
Since the underlying amplitudes associated with dark-matter decay processes are also generally
related through crossing symmetries 
to the amplitudes associated with dark-matter production, scattering, and 
annihilation, the web of dark-matter complementarity relations in multi-component
dark-matter scenarios is significantly enhanced relative to that which emerges
in single-component scenarios. 

In multi-component dark-matter scenarios,
both diagonal and off-diagonal processes in principle contribute
to the signal-event rates at direct- and indirect-detection 
experiments and at colliders.
Indeed, both diagonal and off-diagonal processes are usually permitted by 
the symmetries of the theory.  
However, it turns out that 
the off-diagonal processes generically play only a subleading role 
in the phenomenology of such scenarios.  
There are a variety of reasons for this.
At direct-detection experiments, for example, the rates for up-scattering processes  
(\ie, inelastic processes in which a lighter dark-matter particle scatters into a 
heavier dark-matter particle) are suppressed due to the scattering kinematics.  On the 
other hand, down-scattering processes (\ie, so-called ``exothermic'' processes in which 
a heavier dark-matter particle scatters into a lighter dark-matter particle) frequently 
yield a nuclear recoil energy sufficiently large that the corresponding events are  
vetoed by cuts imposed to reduce experimental backgrounds.  
Likewise, at indirect-detection
experiments, the contribution to the event rate from a co-annihilation process
is proportional to the energy density of each of the two dark-matter 
species involved.  Thus, in cases in which the primordial abundance of one of the
two species is significantly depleted by decays prior to the present epoch, 
this co-annihilation contribution is generically suppressed.  
Finally, in many multi-component
dark-matter scenarios, the mixing among different dark-matter species is the result
of the controlled breaking of approximate symmetries.  In such scenarios, the 
Lagrangian couplings associated with off-diagonal processes are often suppressed 
relative to those associated with the corresponding diagonal processes, 
independent of any kinematic or cosmological considerations. 

In such models, then, it is the diagonal processes which tend to provide the dominant 
contribution to the observed event rates at relevant experiments.
This is unfortunate, since this dominance of the diagonal processes tends to mask
precisely those features which are most 
sensitive to the multi-component nature of the dark sector.

For this reason, discovery and exploration of a potential multi-component dark
sector would be greatly facilitated if such a sector somehow 
incorporated a method of forbidding or suppressing the 
contributions from diagonal processes.  This would allow off-diagonal processes
to shine through and have a significant impact on the resulting phenomenology.
Fortunately, as we shall see in this paper, 
there exist many multi-component dark-matter scenarios in which
this is precisely what occurs --- scenarios in which
the diagonal processes that would otherwise provide the dominant contributions to 
relevant experimental event rates are forbidden or suppressed.  
In such scenarios, the corresponding off-diagonal processes then effectively  
dictate the phenomenology of the dark sector, and even give  
rise to a distinct, enhanced picture of dark-matter
complementarity.

This paper is organized as follows.  In Sect.~\ref{sec:GeneralFeatures}, we
discuss the circumstances under which off-diagonal processes provide the dominant
contribution to event rates at direct-detection experiments, at indirect-detection
experiments, and at colliders.  
We also present a concrete example of a scenario
in which this situation naturally arises.  In Sect.~\ref{sec:RatesCrossSections}, 
we then discuss the phenomenological implications of this dominance of off-diagonal processes
and evaluate the
cross-sections and decay widths for the relevant physical processes.  
In Sect.~\ref{sec:Constraints}, we then focus on the experimental and observational 
considerations which constrain the parameter space of our off-diagonal
dark-matter scenario,   and in   Sect.~\ref{sec:Results} we examine the combined constraints 
on this parameter space and assess the extent to which future 
dark-matter experiments will be able to probe its currently 
unconstrained regions.  
Finally, in Sect.~\ref{sec:Conclusions}, 
we discuss the implications of our results and possible
directions for future work.



\section{Off-Diagonal Interactions\label{sec:GeneralFeatures}}


As discussed in the Introduction, we are interested in exploring the phenomenology
of dark-matter scenarios in which off-diagonal processes provide the dominant 
contribution to event rates at dark-matter detection experiments.
We shall now provide an explicit model in which this is precisely what occurs.

For concreteness, we shall focus on the case in which the dark-matter particles $\chi_i$
are spin-$1/2$ fermions.  We shall also focus on the regime in which the 
leading interactions between the dark and visible sectors at low energies can be modeled 
by a set of contact operators which are separately invariant under charge-conjugation $C$, 
parity $P$, and time-reversal $T$.  For simplicity, we shall focus on the case in which 
the $\chi_i$ couple to the SM fields primarily via dimension-six operators of the form   
\begin{equation}
  \mathcal{O}_{ijq}^{(\alpha)} ~=~ \frac{c_{ijq}^{(\alpha)}}{\Lambda^2} 
     \big[\bar{\chi}_i\Gamma^{(\alpha)}\chi_j\big]
    \big[\bar{q}\Gamma^{(\alpha)} q\big]
      ~,
  \label{eq:D6OperatorForm}
\end{equation}
where $q$ denotes a SM quark,
where $\Lambda$ denotes the scale of new physics, and where $c_{ijq}^{(\alpha)}$ are 
dimensionless coupling coefficients.  
The label $\alpha$ indicates the gamma-matrix 
structure of the fermion bilinears, with $\alpha=\{S,P,V,A,T\}$ corresponding to 
$\Gamma^{(\alpha)}=\{1,\gamma^5,\gamma^\mu,\gamma^\mu\gamma^5,\sigma^{\mu\nu}\}$,
respectively.  Operators with $i=j$ give rise to ``diagonal'' processes involving
two of the same dark-sector particle, while operators with $i\neq j$ give rise to 
``off-diagonal'' processes involving two different dark-sector particles. 

As discussed in the Introduction, when operators with $i=j$ and operators with 
$i\neq j$ are both present with similar values of $c_{ijq}^{(\alpha)}$, the operators 
with $i=j$ tend to play a dominant role in dark-matter phenomenology.  In this paper,   
by contrast, we are primarily interested in studying the alternative possibility in which off-diagonal 
processes are dominant.  There are several ways in which this naturally can occur.  One observation
which we shall exploit in this paper is that
if $\chi_i$ are Majorana rather than Dirac fermions, 
the vector and antisymmetric tensor operators $\mathcal{O}_{ijq}^{(V)}$ and
$\mathcal{O}_{ijq}^{(T)}$ both vanish identically when $i=j$.  Thus, in cases in which 
the primary coupling between such Majorana $\chi_i$ and the visible sector occurs through 
such operators, off-diagonal processes indeed play the dominant role in the resulting 
phenomenology.   

In order to develop this model more fully, let us assume that the dark sector includes a 
vector-like Dirac fermion $\chi$, which comprises a left-handed Weyl spinor 
$\chi_{L\alpha}$ and a right-handed conjugate Weyl spinor 
$\chi_R^{\dagger\dot{\alpha}}$, as well as a complex scalar $\zeta$.  We shall
assume that the action for these dark-sector fields is invariant under an additional
$U(1)'$ gauge symmetry, as well as the discrete symmetries $C$, $P$, and $T$.
Moreover, we shall assume that the $U(1)'$ charges $Q'_\chi$ and $Q'_\zeta$ for 
these fields are chosen such that $Q'_\zeta = - 2Q'_\chi$.  Such a charge assignment 
permits a Yukawa-type interaction of the form
\begin{equation}
  \mathcal{L}_{\mathrm{Yuk}} ~=~ -y\zeta\bar{\chi}\chi^c + \mbox{h.c.}~,
  \label{eq:LYuk} 
\end{equation} 
where $y$ is a real, dimensionless Yukawa coupling and where $\chi^c \equiv C^{-1}\chi C$ 
is the charge-conjugate of $\chi$.  
Note that all of the operators given
in Eq.~(\ref{eq:D6OperatorForm}) with $\chi_i = \chi_j = \chi$ are consistent with 
the symmetries of the theory as well.  

If the scalar $\zeta$ acquires a vacuum expectation 
value (VEV) as a result of some additional dynamics, this VEV breaks $U(1)'$ 
and generates a Majorana mass $m_M = 2 y \langle \zeta\rangle$ for $\chi$ while 
leaving $C$, $P$, and $T$ intact.  Moreover, since $\chi$ is vector-like, a Dirac mass 
$m_D$ for $\chi$ is also consistent with all symmetries of the theory and is therefore
generically expected to be present as well.  Thus, once $\zeta$ acquires a VEV, the mass 
matrix for the Weyl spinors $\chi_{L\alpha}$ and $\chi_{R\alpha}$ generically takes the 
form    
\begin{equation}
  \mathcal{L}_{\mathrm{mass}} ~=~ -\frac{1}{2}\left(\chi_{L}^\alpha,\chi_{R}^\alpha\right)
    \left(\!\begin{array}{cc} m_M & m_D \\ m_D & m_M \end{array} \!\right)\!
    \left(\!\begin{array}{c} \chi_{L\alpha} \\ \chi_{R\alpha} \end{array}\!\!\right) 
    +\, \mbox{h.c.}~~
\end{equation}        
The mass eigenstates of the theory, obtained by diagonalizing this matrix, can therefore be
viewed as a pair of Majorana fermions~\cite{InelasticDM}
\begin{eqnarray}
  \chi_1 &=& \frac{i}{\sqrt{2}}\left(\!\begin{array}{c}
      \chi_{L\alpha} - \chi_{R\alpha}  \\ 
      \chi_L^{\dagger \dot{\alpha}} - \chi_R^{\dagger \dot{\alpha}}
    \end{array}\!\right) \nonumber \\
  \chi_2 &=& \frac{1}{\sqrt{2}}\left(\!\begin{array}{c}
      \chi_{L\alpha} + \chi_{R\alpha}  \\ 
      \chi_L^{\dagger \dot{\alpha}} + \chi_R^{\dagger \dot{\alpha}}
    \end{array}\!\right)~,    
\label{eigenstates}
\end{eqnarray} 
with respective masses $m_1 = m_D - m_M$ and $m_2 = m_D + m_M$.  Moreover, since
$m_M$ is generated dynamically, the mass splitting $\Delta m \equiv m_2 - m_1 = 2m_M$
between these two mass eigenstates can naturally be small.  

Let us assume, for the moment, that amongst the operators
in Eq.~(\ref{eq:D6OperatorForm}),
the vector and antisymmetric-tensor operators dominate.
In such a case,  the full interaction 
Lagrangian ${\cal L}_{\rm int}$ between the dark and visible sectors 
receives contributions from the two operators 
\begin{eqnarray}
  \mathcal{L}_{\mathrm{int}}^{(V)}  &=& \sum_q \frac{2c_q^{(V)}}{\Lambda^2} 
      \big(\bar{\chi}\gamma^\mu\chi\big)\big(\bar{q}\gamma_\mu q\big)~, \nonumber \\
  \mathcal{L}_{\mathrm{int}}^{(T)}  &=& \sum_q \frac{2c_q^{(T)}}{\Lambda^2} 
      \big(\bar{\chi}\sigma^{\mu\nu}\chi\big)\big(\bar{q}\sigma_{\mu\nu}q\big)~.
  \label{eq:LintVecTensRaw}
\end{eqnarray}  
However, when these operators are expressed in
terms of the mass eigenstates in Eq.~(\ref{eigenstates}), the diagonal terms 
vanish~\cite{InelasticDM,KoppSchwetzZupan}. 
The resulting interactions between the dark
and visible sectors are therefore purely off-diagonal, as desired:
\begin{eqnarray}
  \mathcal{L}_{\mathrm{int}}^{(V)}  &=& \sum_q\left[\frac{ic_{q}^{(V)}}{\Lambda^2} 
    \left(\bar{\chi}_1 \gamma^\mu \chi_2 - \bar{\chi}_2 \gamma^\mu \chi_1 \right) 
    \left(\bar{q} \gamma_\mu q \right)\right] ~, \nonumber\\
  \mathcal{L}_{\mathrm{int}}^{(T)}  &=& \sum_q\left[ \frac{ic_{q}^{(T)}}{\Lambda^2} 
    \left(\bar{\chi}_1 \sigma^{\mu\nu} \chi_2 - 
    \bar{\chi}_2 \sigma^{\mu\nu} \chi_1 \right) 
    \left( \bar{q} \sigma_{\mu\nu} q \right)\right]~.~~~~~~~~
\label{eq:LintVecTensMassEigs}
\end{eqnarray}
Indeed, these are the two interaction terms whose phenomenological effects we shall study in the rest
of this paper.
 
Of course, the success of this scenario is predicated on the assumption that only the vector and tensor
interactions dominate from amongst all the operators in
Eq.~(\ref{eq:D6OperatorForm}).
If we had assumed non-negligible coefficients for any of the other operators in
Eq.~(\ref{eq:D6OperatorForm}) and expressed such operators in terms of our mass eigenstates,
both diagonal and off-diagonal operator interactions would have appeared.
However, the operators in
Eq.~(\ref{eq:D6OperatorForm}) 
are only effective operators valid for energies below $\Lambda$.
Thus the question of which operators actually appear
in our effective low-energy theory below $\Lambda$ depends critically on the physics we assume to exist
at higher energy scales above $\Lambda$.

It is not hard to demonstrate that there exist
scenarios in which only the vector and/or tensor operators can be generated in the effective
theory below $\Lambda$.   For example, let us imagine that
the interaction between the dark-sector fermions and the SM quarks arises
due to integrating out the massive gauge boson $Z'^{\mu}$ associated with the 
$U(1)'$ symmetry discussed above.  In particular, let us take the Lagrangian for the dark-sector
fields $\chi$, $\zeta$, and $Z'^{\mu}$ to be 
\beqn
  \mathcal{L} &=&  i\bar{\chi}\Dslash \chi + D^\mu\zeta^\dagger D_\mu\zeta 
    - V(\zeta,\zeta^\dagger) - m_D\bar{\chi}\chi \nonumber\\ 
&& ~~~~~~~~~ - y\zeta \bar{\chi}\chi^c 
    - y\zeta^\dagger \overline{\chi^c}\chi - \frac{1}{4} F'_{\mu\nu}F'^{\mu\nu}~.~~~~~
\label{eq:UrLagrangian} 
\eeqn
Here $F'^{\mu\nu} \equiv (\partial^\mu Z'^\nu - \partial^{\nu}Z'^{\mu})$ is the 
field-strength tensor for $Z'^{\mu}$, while $y$ is a real Yukawa coupling, 
$V(\zeta,\zeta^\dagger)$ is the scalar potential for $\zeta$ and its Hermitian 
conjugate $\zeta^\dagger$, and the covariant derivatives $D_\mu$ are given 
by $D_\mu\chi \equiv \partial_\mu\chi - ig'Q_\chi Z \chi$ and 
$D_\mu\zeta \equiv \partial_\mu\zeta - ig'Q_\zeta Z \zeta$, where $g'$ is the
gauge coupling constant for the $U(1)'$ interaction.  
As discussed above, we assume
that $V(\zeta,\zeta^\dagger)$ is such that $\zeta$ acquires a VEV $\langle\zeta\rangle$
which in turn spontaneously breaks $U(1)'$ and gives a mass $M_{Z'}$ to $Z'^{\mu}$.
If the SM quarks are also charged under $U(1)'$, we may integrate out $Z'^{\mu}$ 
at scales well below $M_{Z'}$ in order to obtain a set of effective operators
coupling $\chi$ to the SM quarks.  In particular, the resulting effective Lagrangian
contains the terms  
\begin{equation}
  \mathcal{L}_{\mathrm{eff}} ~\ni~ \sum_q \frac{g'^2Q_qQ_\chi}{M_{Z'}^2} 
    \big(\bar{\chi}\gamma^\mu\chi\big)\big(\bar{q}\gamma_\mu q\big)~,
\end{equation}
where $Q_q$ is the $U(1)'$ charge of quark $q$.  Moreover, when $\zeta$ acquires
a VEV, a Majorana mass $m_M = 2y\langle\zeta\rangle$ is generated for the dark-sector
fermions, as discussed above.  

Most importantly, however, we see that 
upon converting to the mass eigenbasis
the resulting operators 
in Eq.~(\ref{eq:UrLagrangian}) 
take the vector-operator form appearing in Eq.~(\ref{eq:LintVecTensMassEigs}).
Indeed, we can now identify $\Lambda$
in terms of the parameters of our underlying theory via
$c_q^{(V)}/\Lambda^2 \sim g'^2 Q_q Q_\chi /M_{Z'}^2$.
Moreover, no other operator with a different Lorentz structure appears
below $\Lambda$ at leading order. 

Of course, it is always possible to add additional interaction terms in our ultraviolet (UV) 
theory in Eq.~(\ref{eq:UrLagrangian})  
in order to generate the full spectrum of effective operators in 
Eq.~(\ref{eq:D6OperatorForm}) at low energies.  Thus the question of what is fully ``natural''
becomes a question of presupposing a particular UV theory --- a task which is beyond the scope
of this paper.   We have nevertheless demonstrated that the model presented in this section gives
rise to off-diagonal interactions while at the same time suppressing diagonal interactions ---
all emerging in a unified way from the assumption of a simple Lorentz structure for our dimension-six
effective interactions between the dark and visible sectors.


\section{Rates and Cross Sections\label{sec:RatesCrossSections}}

  
Off-diagonal dark-matter scenarios exhibit a rich set of 
complementarity relations~\cite{DecayComplementarity}. 
Indeed, as discussed in the Introduction and in Ref.~\cite{DecayComplementarity}, 
a single off-diagonal operator can simultaneously
give rise to inelastic up-scattering and down-scattering processes at direct-detection 
experiments, dark-matter co-annihilation processes relevant for indirect detection, 
asymmetric dark-matter production at colliders, and decay processes in which a heavier
dark-matter particle decays into a lighter one.  In this section, we evaluate the 
cross-sections and decay rates for these processes.

We shall perform these calculations within the framework of the model introduced
in Sect.~\ref{sec:GeneralFeatures}.   Specifically, we shall imagine that
the dark sector comprises two Majorana fermions $\chi_1$ and 
$\chi_2$ with masses $m_1$ and $m_2$  which couple to the SM quarks
through either of the operators in Eq.~(\ref{eq:LintVecTensMassEigs}).
Without loss of generality we shall take $m_2 > m_1$ 
so that the mass splitting $\Delta m \equiv m_2 - m_1$ is positive-definite.  In
what follows, we shall also focus on the regime in which 
$\Delta m \lesssim \mathcal{O}(\mbox{MeV}) \ll m_1 \approx m_2$, as this is the regime 
in which inelastic processes have a demonstrable impact on direct-detection phenomenology
and in which $\chi_2$ can be sufficiently long-lived as to have observable 
consequences for indirect detection.  In other words, this is the regime in which 
the off-diagonal nature of the dark-sector interactions truly matters. 
Finally, 
we shall assume that 
the dark-sector fields only couple to first-generation quarks --- \ie, we shall take 
$c_q^{(V)} = c_q^{(T)} = 0$ for $q \in \{c,s,b,t\}$ within the operators 
in Eq.~(\ref{eq:LintVecTensMassEigs}).
We shall then calculate our cross-sections and decay rates for these two operators respectively.

Of course, the direct- and indirect-detection phenomenology of our scenario depends not only on the 
particle-physics properties of our dark-sector fields, but also on their astrophysical
properties --- and in particular, on their cosmological abundances.  
We shall therefore assume that the contributions from $\chi_1$ and $\chi_2$ together 
constitute essentially the entire dark-matter abundance 
$\OmegaDM\approx 0.26$~\cite{Planck} at present time $\tnow$.
We likewise assume that initial number densities $n_1(t_0)$ and $n_2(t_0)$ are established at 
some early time $t_0 \ll \tnow$, and that the only subsequent change in these number
densities, other than the usual dilution resulting from Hubble expansion,
is due to dark-matter co-annihilation or decay processes following from one of 
the operators in Eq.~(\ref{eq:LintVecTensMassEigs}).  Since the co-annihilation 
rate at time $t$ is proportional to the product $n_1(t)n_2(t)$ while the depletion
rate of $\chi_2$ due to decays is proportional to $n_2(t)$ alone, it is sufficient to 
focus on decays.  
For the operators in Eq.~(\ref{eq:LintVecTensMassEigs}),
each decay of a $\chi_2$ particle produces a $\chi_1$ particle
and thus the total comoving number density $n_{\mathrm{tot}}(t) = n_1(t) + n_2(t)$ 
of dark-matter particles is 
``conserved'' in the sense that its only time-dependence comes from Hubble expansion rather than
net particle creation/annihilation.
  Therefore, if we define the initial fractions 
$f_i(t) \equiv n_i(t)/n_{\mathrm{tot}}(t)$ of the dark-matter number density 
contributed by $\chi_i$ at any time $t$, we have at present time
\begin{eqnarray}
  f_2(\tnow) &=& f_2(t_0) \, e^{-(\tnow-t_0)/\tau_2} \nonumber \\
  f_1(\tnow) &=& 1 - f_2(\tnow)~,
  \label{eq:DMFracsNow}
\end{eqnarray}
where $\tau_2$ is the lifetime of $\chi_2$.  In particular, all effects due to Hubble
expansion are eliminated from the $f_i$.   Moreover, since 
we have assumed that $\Delta m \ll m_1 \approx m_2$, we see that
$f_1(\tnow)$ and $f_2(\tnow)$ are to a very good 
approximation equal to the fractions of $\OmegaDM$ contributed by the respective 
dark-matter particles.            

In what follows, we shall take the {\it primordial}\/ number-density fraction 
$f_2(t_0)$ to be the free parameter which characterizes the relative abundances 
or number densities of the dark-matter particles. 
Moreover, in order to distinguish between the vector or tensor cases
[\ie, in order to distinguish between the two interaction Lagrangians in Eq.~(\ref{eq:LintVecTensMassEigs})],
we shall adopt the notation    $f_{1,2}^{(V,A)} \equiv f_{1,2}(t_0)$ 
for these two cases, respectively.

\subsection{Decay Rates \label{sec:DecayRates}}

As stressed in Ref.~\cite{DecayComplementarity},
one of the most important and unique consequences of off-diagonal interactions among the dark-sector
fields is the possibility for a heavier dark-matter particle to decay into a lighter
dark-matter particle plus additional SM fields.  In the regime in which  
$\Delta m \lesssim 1~\mev$, the only SM particles which can appear in the 
final state are photons and neutrinos.  Since decay processes involving neutrinos are 
generically suppressed relative to those involving photons alone, we focus exclusively 
on the latter.  

The operators in Eq.~(\ref{eq:LintVecTensMassEigs}) which describe the microscopic
interactions between our dark-sector particles $\chi_i$ and SM quarks ultimately give 
rise to effective interactions between $\chi_1$, $\chi_2$, and the photon field in the 
low-energy, macroscopic theory.  The structure of the corresponding effective operators 
can be computed, for example, within the framework of chiral perturbation theory.  
However, the structure of the leading operators can also be 
inferred simply from symmetry considerations.  For example, we note that both 
the vector-current density $\chi_i\gamma^\mu\chi_j$  and the tensor-current 
density $\chi_i\sigma^{\mu\nu}\chi_j$ are odd under charge conjugation.
Since the photon field is likewise $C$-odd, each of these operators can only couple 
to an odd number of photons.  Thus, na\"{i}vely, one would expect that the leading
contribution to the decay width of $\chi_2$ would arise due to the two-body process     
$\chi_2 \rightarrow \chi_1 \gamma$.  

However, for dark-sector particles which couple to the visible sector primarily 
through the vector operator in Eq.~(\ref{eq:LintVecTensMassEigs}), there is an
additional consideration which one must take into account.  Specifically, the Ward identity
prohibits the coupling of a current operator of the form $\bar{\chi}_i\gamma^\mu\chi_j$ with 
$i\neq j$ to a single photon.  The leading contribution to the decay width of $\chi_2$
is therefore associated with the four-body decay process 
$\chi_2 \rightarrow \chi_1 \gamma \gamma \gamma$.  
Contributions to this process can arise both from contact interactions involving the 
dark-sector fields and the photon field alone and from processes involving a photon and an 
off-shell $\pi^0$ which subsequently decays into a pair of photons.  We find that the leading contribution is of the latter type.  The corresponding operator in the effective low-energy 
Lagrangian is a consequence the chiral anomaly, and therefore the operator coefficient may be 
computed exactly from the Wess-Zumino-Witten term~\cite{WessZumino,WittenCurrentAlgebra}.  
The result is
\begin{eqnarray}
  \mathcal{O}_{\mathrm{eff}}^{(V)} &=& -\frac{i\big[2c_u^{(V)} + c_d^{(V)}\big]e}
    {16\pi^2 f_\pi \Lambda^2} \, \epsilon_{\mu\nu\rho\sigma} \nonumber \\
    && ~~~\times  \big(\bar{\chi}_1 \gamma^\mu \chi_2 - \bar{\chi}_2 \gamma^\mu \chi_1\big)
     F^{\nu\rho}(\partial^\sigma\pi^0)~,~~~~~
  \label{eq:OpDecayToPhotonsV}
\end{eqnarray}
where $e$ is the absolute value of the electron charge, $f_\pi$ is the pion-decay constant, and 
$F_{\mu\nu}$ is the photon field-strength tensor.

By contrast, for dark-sector particles which couple to the visible sector primarily 
through the tensor operators in Eq.~(\ref{eq:LintVecTensMassEigs}), no additional
considerations forbid a coupling of the corresponding current operator
$\bar{\chi}_i\sigma^{\mu\nu}\chi_j$ to a single photon.  Thus, in this case, the 
leading contribution to the decay width of $\chi_2$ is indeed associated with the 
processes $\chi_2 \rightarrow \chi_1 \gamma$. 
The operator in the effective low-energy Lagrangian which provides the leading contribution 
to this process can be written in the form
\begin{eqnarray}
  \mathcal{O}_{\mathrm{eff}}^{(T)} &\equiv & 
     \frac{i\Lambda_\mathrm{QCD}}{\Lambda^2}
     \left[\xi_uc_u^{(T)} + \xi_dc_d^{(T)}\right] \nonumber \\
    & & \times \big(\bar{\chi}_1 \sigma^{\mu \nu} \chi_2 - 
    \bar{\chi}_2 \sigma^{\mu \nu} \chi_1 \big) F_{\mu \nu}~,
  \label{eq:OpDecayToPhotonsT}
\end{eqnarray}
where $\Lambda_{\mathrm{QCD}}$ is the QCD scale and where $\xi_u$ and $\xi_d$ are
dimensionless coefficients which parametrize our ignorance of the underlying strong 
dynamics.  Unlike the coefficients in Eq.~(\ref{eq:OpDecayToPhotonsV}), 
$\xi_u$ and $\xi_d$ are not directly calculable from first principles.  
Nevertheless, is reasonable to expect that these coefficients are $\mathcal{O}(1)$. 

Given the operators in Eqs.~(\ref{eq:OpDecayToPhotonsV}) and~(\ref{eq:OpDecayToPhotonsT}), 
the corresponding contributions to the $\chi_2$ decay widths are given by
\begin{eqnarray}
  \Gamma_{\chi_2 \rightarrow \chi_1 \gamma \gamma \gamma}^{(V)} &=& 
    \frac{\big[2c_u^{(V)} + c_d^{(V)}\big]^2\alpha^3}{25025 \cdot 3^4\cdot 2^5\cdot\pi^{10}}
    \frac{(\Delta m)^{13}}{f_\pi^4 m_\pi^4 \Lambda^4}~,\nonumber\\
  \Gamma_{\chi_2 \rightarrow \chi_1 \gamma}^{(T)} &=& 
    \frac{4\Lambda_{\mathrm{QCD}}^2 (\Delta m)^3}{\pi \Lambda^4}
    \big[\xi_uc_u^{(T)} + \xi_dc_d^{(T)}\big]^2~.~~~~~~~~~
\label{eq:DecayRates}
\end{eqnarray}

\subsection{Co-annihilation Cross-Sections\label{sec:CoannXSecs}}

In addition to decays, the operators in Eq.~(\ref{eq:LintVecTensMassEigs}) also 
give rise to co-annihilation processes of the form $\chi_1\chi_2 \rightarrow \bar{q}q$.
Such co-annihilation processes can also be relevant for indirect detection.  
The corresponding matrix elements can be found, \eg, in 
Ref.~\cite{Kumarfatia}.  In both the vector and antisymmetric-tensor cases, we
find that the cross-section is $s$-wave even in the $\Delta m \rightarrow 0$ limit.  
Thus, for $\Delta m \ll m_1 \approx m_2$, the thermally averaged cross-sections for 
dark-matter co-annihilation in galactic halos are independent of $\Delta m$ at leading 
order and are given by
\begin{eqnarray}
  \langle \sigma v \rangle_{\chi_1 \chi_2}^{(V)} &=& 
    \frac{3 m_1^2}{\pi \Lambda^4} \sum_q \left[c_q^{(V)}\right]^2~, \nonumber\\
  \langle \sigma v \rangle_{\chi_1 \chi_2}^{(T)} &=& 
    \frac{6 m_1^2}{\pi \Lambda^4} \sum_q \left[c_q^{(T)}\right]^2~.
  \label{eq:CoannXSecs}
\end{eqnarray}
Of course, these results reflect the contact-operator coupling between the dark and visible sectors.  
By contrast, in the regime in which these sectors are coupled by a sufficiently light mediator, dark-matter co-annihilation can be predominantly $p$-wave at late times, owing to Sommerfeld enhancement~\cite{Das}.

It is important to note that the expressions in Eq.~(\ref{eq:CoannXSecs})
assume that processes of the form $\chi_1\chi_2 \rightarrow \bar{q}q$
provide the dominant contribution to the dark-matter co-annihilation 
cross-section and that the effective-theory description of the interactions
between the dark and visible sectors in Eq.~(\ref{eq:LintVecTensMassEigs}) 
remains valid up to the energy scales $\sqrt{s} \sim m_1 + m_2$ relevant 
for the co-annihilation of a population of non-relativistic dark-matter particles.
By contrast, if the effective theory breaks down at lower energies, other processes 
may dominate the co-annihilation cross-section.  For example, if the contact operators in 
Eqs.~(\ref{eq:LintVecTensRaw}) and (\ref{eq:LintVecTensMassEigs}) 
arise in the low-energy effective 
theory due to the presence of a massive vector mediator $\phi$ of mass $m_\phi$ in the 
UV theory, dark-matter co-annihilation to a pair of on-shell $\phi$ 
particles typically dominate the thermally averaged cross-section when 
$m_\phi \lesssim m_1 \approx m_2$.  Moreover, since the antisymmetric-tensor 
operator in Eqs.~(\ref{eq:LintVecTensRaw}) and (\ref{eq:LintVecTensMassEigs}) 
does not respect the full electroweak 
gauge symmetry of the SM, such an operator should be viewed as an effective 
operator which arises at low energies as a consequence of the electroweak symmetry breaking 
generated by the non-zero VEV of the SM Higgs field. 
Thus, one might expect additional annihilation channels involving 
the Higgs boson to open up for dark-matter masses 
$m_1 + m_2 \gtrsim m_h \approx 125$~GeV.~  Such processes can in principle also
contribute significantly to the dark-matter co-annihilation rate.    

Clearly, the contributions to $\langle \sigma v \rangle_{\chi_1 \chi_2}^{(V,T)}$ 
from additional processes such as those discussed above are highly model-dependent.  
For sake of generality, we therefore refrain from specifying a particular UV 
completion for the effective operators in Eqs.~(\ref{eq:LintVecTensRaw}) and (\ref{eq:LintVecTensMassEigs}) 
in this analysis, and instead focus on the regime in which the expressions in Eq.~(\ref{eq:CoannXSecs}) provide
an accurate description of the thermally-averaged co-annihilation cross-sections for the 
vector and antisymmetric-tensor cases.

\subsection{Differential Cross-Sections for Inelastic Scattering\label{sec:DirectDetection}}

The matrix element associated with dark-matter scattering is independent of 
$\Delta m$ at lowest order.  As a result, the main dependence of the differential scattering rate 
on $\Delta m$ arises from the phase space.  

In general, the differential cross-section for a dark-matter particle $\chi$ of 
mass $m_\chi$ scattering off a target nucleus of mass $m_A$ can be written in the form
\begin{equation}
  \frac{d\sigma^{(V,T)}}{dE_R} ~=~ \frac{m_A}{2 \mu_{\chi A}^2 v^2}\, 
    \sigma_{0A}^{(V,T)} \, F_{V,T}^2 (E_R)~,   
  \label{eq:diffsigma}
\end{equation}
where $E_R$ is the recoil energy of scattered nucleus in the detector frame,
where $v$ is the detector-frame velocity of $\chi_i$, 
where $\mu_{\chi A} \equiv  m_\chi m_A /(m_\chi + m_A )$ 
is the reduced mass of the $\chi$-nucleus system, where 
$F^{(V,T)}(E_R)$ is the appropriate nuclear form factor, and where $\sigma_{0A}^{(V,T)}$ is 
the scattering cross-section at zero momentum transfer.  We observe that the vector 
interaction in Eq.~(\ref{eq:LintVecTensMassEigs}) contributes to spin-independent (SI) 
scattering, while the antisymmetric-tensor interaction only contributes 
to spin-dependent (SD) scattering.  Thus, we adopt a parametrization for $\sigma_{0A}^{(V)}$ 
and $\sigma_{0A}^{(T)}$ of the form 
\begin{eqnarray}
  \sigma_{0A}^{(V)} &=& \frac{4\mu_{\chi A}^2}{\pi\Lambda^4} 
    \Big[Z B_p^{(V)} \!+ (A-Z) B_n^{(V)} \Big]^2 ~,\nonumber\\
  \sigma_{0A}^{(T)} &=& \frac{16\mu_{\chi A}^2}{\pi\Lambda^4} 
    {J_A + 1 \over J_A} \Big[\langle S_p \rangle B_p^{(T)} \!+ 
    \langle S_n \rangle B_n^{(T)} \Big]^2~.~~~~~
\end{eqnarray}
Here $Z$ and $A$ respectively denote the numbers of protons and total nucleons
in the target nucleus while $J_A$ is the total nuclear spin and 
$\langle S_N \rangle$ with $N\in\{p,n\}$ represents the average spin projection of 
the corresponding nucleon $N$ within the nucleus.  
The dimensionless couplings $B^{(V,T)}_{N}$ are given by
\begin{equation}
  B^{(V,T)}_{N} ~=~ \sum_q c_q^{(V,T)} \, \Delta q^{(V,T)}_{N} 
    ~ \Xi_q^{(V,T)}(m_\chi,\mu_N)~,~~~
\label{Xicontaining}
\end{equation}
where the effective nucleon form factors $\Delta q^{(V,T)}_{N}$ 
have the values~\cite{BhattacharyaLattice}
\begin{eqnarray}
    \Delta u^{(V)}_{p} &=~ \Delta d^{(V)}_{n} &=~ 2 \nonumber \\
    \Delta d^{(V)}_{p} &=~ \Delta u^{(V)}_{n} &=~ 1 \nonumber \\
    \Delta s^{(V)}_{p} &=~ \Delta s^{(V)}_{n} &=~ 0 \nonumber \\
    \Delta u^{(T)}_{p} &=~ \Delta d^{(T)}_{n} &=~ 0.774 \nonumber \\
    \Delta d^{(T)}_{p} &=~ \Delta u^{(T)}_{n} &=~ -0.223 \nonumber \\
    \Delta s^{(T)}_{p} &=~ \Delta s^{(T)}_{n} &=~ 0.008~.
\label{eq:VecTensCharges}
\end{eqnarray}
The factor $\Xi_q^{(V,T)}(m_\chi,\mu_N)$ in Eq.~(\ref{Xicontaining}) 
accounts for the renormalization-group 
evolution of the effective contact operator from the energy scale $m$ 
of the dark-matter particle down to the nucleon scale $\mu_N \sim 1 - 2$~GeV.~
For the vector case, gauge invariance implies that $\Xi^{(V)}(m_\chi,\mu_N)=1$.  By
contrast, for the antisymmetric-tensor case, $\Xi^{(T)}(m_\chi,\mu_N)$ is a product of
factors of the form~\cite{HillSolon}   
\begin{eqnarray}
  X_q^{(T)}(\mu_i,\mu_j) &=& \frac{m_q (\mu_j)}{m_q (\mu_i)}
    \left[\frac{\alpha_s (\mu_j)}{\alpha_s (\mu_i)} \right]^{-16/3\beta_0(\mu_j)}
    \nonumber\\ 
      && ~~~~\times \big[1 + \mathcal{O}(\alpha_s) \big]~,
\end{eqnarray}
where $m_q$ and $\alpha_s$ are the running quark mass and QCD coupling  
in the $\overline{\rm MS}$ renormalization scheme,
and where $\beta_0(\mu_j) = 11 - (2/3) n_f(\mu_j)$ is the beta function, 
which depends on on the number $n_f(\mu_j)$ 
of quark flavors $q$ with masses $m_q > \mu_j$.  In particular, for 
\mbox{$m_\chi < m_b$} we have
{$\Xi_q^{(T)}(m_\chi,\mu_N) = X_q^{(T)}(m_\chi,\mu_N)$}, while for 
\mbox{$m_b < m_\chi < m_t$} we have
{$\Xi_q^{(T)}(m_\chi,\mu_N) = X_q^{(T)}(m_\chi,m_b)X_q^{(T)}(m_b,m_N)$} 
and for \mbox{$m_\chi > m_t$} we have 
{$\Xi_q^{(T)}(m_\chi,\mu_N) = X_q^{(T)}(m_\chi,m_t)X_q^{(T)}(m_t,m_b)X_q^{(T)}(m_b,\mu_N)$}.   

The differential event rate expected per unit target mass at a given detector
can be computed in a straightforward manner from the differential cross-section in
Eq.~(\ref{eq:diffsigma}).  The result is 
\begin{equation}
  \frac{dR}{dE_R} ~=~ \frac{\rho^{\mathrm{loc}}}{m_A m_\chi} \epsilon(E_R)
    \int_{v > \vmin}\frac{d\sigma^{(V,T)}}{dE_R} \, v \mathcal{F}(\vec{v}) \, d^3v~,~~~
  \label{eq:diffRateRaw}
\end{equation}
where $\rho^{\mathrm{loc}}$ is the local energy density of the dark-matter species
in question, where $\mathcal{F}(\vec{v})$ is the velocity distribution
of that species in the local dark-matter halo, where $\epsilon(E_R)$ is the detector 
efficiency expressed as a function of $E_R$, and where $\vmin$ is the minimum value
of $v\equiv |\vec{v}|$ that is kinematically required in order for scattering to take place.
This differential rate can also be expressed in the more compact form 
\begin{equation}
  \frac{dR}{dE_R} ~=~
    \frac{\rho^{\mathrm{loc}}\sigma_{0}^{(V,T)}}{2 m \mu_{\chi A}^2}  
    \, \epsilon(E_R)\, F^{2}_{V,T}(E_R)\, \mathcal{I}(E_R)~, 
  \label{eq:diffRate}
\end{equation} 
where we have defined $\mathcal{I}(E_R)$ as a shorthand notation for the dimensionless 
integral    
\begin{equation}
  \mathcal{I}(E_R) ~\equiv~ \int_{v>\vmin} \frac{\mathcal{F}(\vec{v})}{v}\, d^3v~. 
  \label{eq:IFunc}
\end{equation}

In our off-diagonal dark-matter scenario, the scattering processes relevant for
direct detection are purely inelastic.  Moreover, the event rate 
generically includes contributions from both the up-scattering of $\chi_1$ particles 
in the local dark-matter halo and the down-scattering of $\chi_2$ particles.
For simplicity, we shall assume that the local energy densities of these two
particles are proportional to their respective cosmological number-density fractions 
$f_1^{(V,T)}(\tnow)$ and $f_2^{(V,T)}(\tnow)$.  We take the total local dark-matter
energy density to be $\rhotot^{\mathrm{loc}} \approx 0.3 \, \gev \, \cm^{-3}$.    
We shall also take the velocity distributions of both $\chi_1$ and $\chi_2$ to be Maxwellian 
in the frame of the dark-matter halo, with a one-dimensional velocity dispersion given by
$v_0/\sqrt{2}$, where $v_0 \approx 220$~km/s is the local circular velocity.  However, this
distribution is truncated above the galactic escape velocity $v_\text{esc}\approx 540$~km/s.  
For a velocity distribution of this form, the integral over detector-frame velocities in 
Eq.~(\ref{eq:IFunc}) can be performed analytically as a function of $\vmin$ 
(see, \eg, Ref.~\cite{SavageFreese}).
  
The total differential event rate in our off-diagonal dark-matter scenario is simply
a sum of the event rates for the up-scattering of $\chi_1$ and the down-scattering of
$\chi_2$.  Since the coupling coefficients for these two processes are equal --- after all, 
they follow from the same Lagrangian operator --- the only difference between the 
corresponding differential rates is due to kinematics.  Indeed, even in 
the regime in which $\Delta m \ll m_1 \approx m_2$, there yet remains one crucial 
difference between up-scattering and down-scattering kinematics:  a difference in the 
dependence of the threshold velocity $\vmin$ on $\Delta m$.  In general, for a 
dark-matter particle scattering inelastically with an atomic nucleus, this threshold 
velocity is given by   
\begin{eqnarray}
  \vmin &\approx & \frac{1}{\sqrt{2 m_A E_R}} 
    \left|\frac{E_R m_A}{\mu_{\chi A}} \pm \Delta m \right|~,
  \label{eq:vmin}
\end{eqnarray}
where the plus (minus) sign corresponds to the case of a dark-matter particle up- (down-)scattering into a 
heavier (lighter) dark-sector state.

The total event rate is the sum of the rates for both up-scattering and down-scattering.
For compactness, since we are working in the regime in which $\Delta m \ll m_1 \approx m_2$, 
we shall henceforth replace $m_1$ and $m_2$ by a single mass parameter $m$ and retain
the dependence on $\Delta m$ in $\vmin$.  For a detector medium made of different nuclei $A$ 
of mass fractions $w_A$, we may express this total rate in the compact form
\begin{eqnarray}
    R &=& \sum_A\frac{\rhotot^{\mathrm{loc}}w_A}{2 m \mu_{\chi A}^2} 
    \, \sigma_{0A}^{(V,T)} 
    \, \Big[f^{(V,T)}_1(\tnow) \mathcal{K}_A^+(m) \nonumber \\ & & ~~~~~~~~+ 
    f^{(V,T)}_2(\tnow) \mathcal{K}_A^-(m)  \Big]~
  \label{eq:rateGeneral}
\end{eqnarray}
by defining
\begin{equation}
  \mathcal{K}_A^\pm (m) ~\equiv~ \int d E_R \, F_{V,T}^2(E_R) 
    \, \epsilon(E_R) \, \mathcal{I}_\pm(E_R)~,
\end{equation}
where $\mathcal{I}_\pm(E_R)$ denotes the integral in Eq.~(\ref{eq:IFunc}) with 
the appropriate choice of $\vmin$ for up-scattering (plus sign) or down-scattering 
(minus sign).

\subsection{Collider Production\label{sec:ColliderRates}} 

The most relevant channels for dark-matter detection at hadron colliders in our
off-diagonal dark-matter scenario involve asymmetric pair-production via the process 
$q\bar{q} \rightarrow \chi_1\chi_2$ in association with one or more additional 
jets, with a photon, or with a $W^\pm$ or $Z$ boson.
For the purposes of our eventual study,
cross-sections for these processes were derived using
MadGraph~5~\cite{MadGraph} with model input from the FeynRules 
package~\cite{FeynRules1,FeynRules2}.  Detector-level event rates 
were obtained using MadGraph~5 for event-generation in conjuction with 
Pythia~6.4~\cite{Pythia} for fragmentation and hadronization 
and Delphes 3.3.0~\cite{Delphes} for detector simulation.


\section{Constraints\label{sec:Constraints}}


The fundamental parameters which govern the phenomenology of our off-diagonal 
dark-matter scenario in the $\Delta m \ll m_1 \approx m_2$ regime are the 
scale $\Lambda$, the mass-splitting parameter $\Delta m$, the mass scale
$m$ of the dark-matter particles, the operator coefficients $c_u^{(V,T)}$ and 
$c_d^{(V,T)}$, and the primordial abundance fraction $f_2^{(V,T)}$.  Moreover, 
in the antisymmetric-tensor case, the decay width of $\chi_2$ depends on 
the unknown $\mathcal{O}(1)$ factors $\xi_u$ and $\xi_d$ appearing in 
Eq.~(\ref{eq:OpDecayToPhotonsT}).  

That said, some of these parameters --- for example, $\xi_u$ and $\xi_d$ and the 
ratio of $c_u^{(V,T)}$ to $c_d^{(V,T)}$ --- have a less significant impact 
on the phenomenology than others.  Thus, in order to assess the extent to which 
current data constrain the parameter space of our scenario --- and the extent to 
which future experiments could potentially probe additional regions of that parameter 
space --- we shall adopt two simplifying assumptions with regard to these 
``secondary'' parameters.  First, we shall take $c_u^{(V,T)} = c_d^{(V,T)}$.  
For such a coupling structure, we may absorb these coefficients into the scale $\Lambda$ 
without further loss of generality through the redefinition 
$\Lambda/c_{q}^{(V,T)} \rightarrow \Lambda$.  In addition, for 
concreteness, we shall also assume values for $\xi_u$ and $\xi_d$ such that
$M_\ast \equiv (\xi_u + \xi_d)\Lambda_{\mathrm{QCD}} = 1$~GeV.~  Again, we stress 
that our results are not particularly sensitive to the precise value of $M_\ast$,
provided that $M_\ast \sim \mathcal{O}(\mbox{GeV})$.  

With these assumptions,
the number of parameters which govern our scenario reduces to four:  
$\Lambda$, $\Delta m$, $m$, and $f_2$. 
Moreover, from amongst these parameters, it is $\Delta m$ which 
encapsulates (and in some
sense quantifies) the off-diagonal nature of our scenario.   
Indeed, the $\Delta m\to 0$ limit of our results
corresponds to the ``diagonal'' limit:   in this limit we find that
$\tau_2$ (the lifetime of the heavier component) becomes infinite
(thereby turning off the decay process), 
up-scattering and down-scattering become identical elastic processes,
and so forth.  Thus our main interest in this paper lies in 
studying how the overall phenomenology of our model shifts as we increase $\Delta m$ from zero.

Before we can proceed, however, we must recognize that
a variety of experimental probes of the dark sector already constrain the parameter
space of our model.  These include indirect-detection probes of both dark-matter
decay and co-annihilation in the galactic halo as well as the results of direct-detection experiments 
and the results of searches for excesses of events in channels with large missing transverse energy 
at the Large Hadron Collider (LHC).  We shall therefore discuss each of these constraints in turn.

\subsection{Decay Constraints\label{sec:DecayConstraints}}
 
When assessing the phenomenological consequences of dark-matter decay in our 
scenario, the most important consideration is how the lifetime $\tau_2$ of the
heavier dark-sector state relates to $\tnow$ within the region of parameter space 
in which we are primarily interested --- \ie, the region in which 
$\Delta m \lesssim \mathcal{O}(\mev)$ and $\Lambda \gtrsim \mathcal{O}(\gev)$.
  
In the vector-interaction case, 
we see from Eq.~(\ref{eq:DecayRates}) that $\tau_2 \propto (\Delta m)^{-13}$, which
implies that $\tau_2 \gtrsim 10^{14}\,\tnow$ within our region of interest.  
Thus, in this case, we may take 
\begin{equation}
  f^{(V)}_2(\tnow) ~\approx~ f^{(V)}_{2}~.
\end{equation}  
Moreover, since $\tau_2$ is far too long for dark-matter decays to contribute 
appreciably to event rates at indirect-detection experiments, decay considerations
place no meaningful bounds on the parameter space of our off-diagonal dark-matter 
scenario.  

By contrast, in the antisymmetric-tensor-interaction case, we have 
\begin{equation}
  \tau_2 ~=~ \frac{\pi \Lambda^4}{4M_\ast^2 (\Delta m)^3}~.
  \label{eq:tau2Tens}
\end{equation}  
Thus, in this case, the range of lifetimes accessible
within our region of interest extends from the very short 
to the cosmologically stable --- \ie, from $\tau_2 \ll \tnow$ to $\tau_2 \gg \tnow$. 
Thus, the effects of dark-matter decay cannot generally be neglected, and the 
present-day number-density fraction of $\chi_2$ in our scenario, under the 
simplifying assumptions discussed above, is given by
\begin{equation}
  f^{(T)}_{2}(\tnow)  ~=~ f_{2}^{(T)} \exp \left[
    -\frac{4M_\ast^2 (\Delta m)^3 t_{\rm now}}{\pi \Lambda^4}\right]~. 
  \label{eq:f2Tens}
\end{equation}

In the antisymmetric-tensor-interaction case, the decays of $\chi_2$ can also 
give rise to observable signals at indirect-detection experiments. 
As discussed in Sect.~\ref{sec:DecayRates}, the primary decay channel for $\chi_2$
in the antisymmetric-tensor case is the two-body process 
$\chi_2 \rightarrow \chi_1 \gamma$.  The resulting primary-photon spectrum from
this process consists of a single monochromatic line at 
$E_\gamma \approx \Delta m \leq \mathcal{O}(\mev)$.
Observational limits on such a photon signal can be derived from searches for 
line-like X-ray signals emanating from sources such as the halo of the 
Milky Way~\cite{BoyarskyMWHalo}, the Andromeda galaxy~\cite{BoyarskyXMMNewton}, 
dwarf spheroidal galaxies~\cite{MalyshevDwarfs,RuchayskiyXMMNewton}, and galaxy 
clusters~\cite{BoyarskyDecayBoundCluster}.  Indeed, the results of such searches are 
used to constrain sterile-neutrino models in which a sterile neutrino of mass $m_{\nu_s}$ 
decays into an active neutrino of mass $m_{\nu_{\ell}}$ and a photon of energy
$E_\gamma \approx E_{\nu_{\ell}} \approx m_{\nu_s} / 2$.  In particular, these results place limits
on the flux of such photons, a quantity which is proportional to the product of the number density 
of the sterile neutrino and its decay rate.  Adapting these limits to off-diagonal 
dark-matter scenario is straightforward.  Indeed, the only difference 
is that the bound on the lifetime is suppressed by an additional factor of 
$2\Delta m / m_2$ because $n_2(\tnow)$ scales inversely with $m_2$ for fixed
abundance.

For a sterile neutrino, one finds that the bound on the corresponding lifetime 
$\tau_{\nu_s}$ is essentially independent of $m_{\nu_s}$ over 
a large range of $m_{\nu_s}$ and is given by 
$\tau_{\nu_s} \gtrsim 10^{27} \, \s$~\cite{BoyarskyRuchayskiy}.  The corresponding 
constraint on the parameter space of our off-diagonal dark-matter scenario is
therefore
\begin{equation}
  f_{2}^{(T)}(\tnow) \Gamma_2^{(T)} ~\lesssim~ \left( \frac{m}{2 \Delta m} \right ) 
    \times 10^{-27} \text{ s}^{-1}~.
  \label{eq:GamTensBound}
\end{equation}
This bound can then be translated into a constraint on the fundamental parameters $\Lambda$, 
$\Delta m$, and $m$ which characterize our off-diagonal dark-matter scenario through the use 
of Eqs.~(\ref{eq:DecayRates}) and (\ref{eq:f2Tens}).

In passing, we also note that if $\Delta m \sim 3.5~\kev$, the resulting 
$E_\gamma \sim 3.5$~keV line from $\chi_2$ decay could potentially explain the excess 
of X-rays observed in galaxy clusters and in the halos of both Andromeda and the Milky Way 
(for recent discussions, see, \eg, Refs.~\cite{QueirozOverview,SterileNeutrinoBonds}).
In order to determine the region for the number-density fraction $f_2$ necessary to explain such a
signal, we may once again proceed by analogy with the case of a decaying sterile neutrino. 
For a particle of this sort with an abundance $\Omega_{\nu_s}\sim \OmegaDM$, a
mass $m_{\nu_s} \sim 7$~keV, and a lifetime in the range 
$\tau_{\nu_s} \sim (2\text{--}20) \times 10^{27}~\s$ are required to account for 
the observed excess.  Likewise, in our off-diagonal dark-matter scenario, we can
account for such an excess provided that $\Delta m = 3.5~\kev$ and that 
\begin{equation}
  f_2^{(T)}(\tnow) \Gamma_2^{(T)} ~\sim ~ (2\text{--}20) \times 
    \left( {m \over 7~\kev } \right) \times 10^{27}\, s^{-1}~. 
  \label{eq:3keVLineBound}
\end{equation} 
However, we note that recent observations of the Perseus cluster by the Hitomi 
satelite~\cite{Hitomi} show no evidence of such a line.  

Finally, an additional constraint on dark-matter decays in our scenario arises due
to the impact these decays can have on the ionization history of the cosmic microwave
background (CMB).~  Indeed, if a significant population of $\chi_2$ particles decay
after recombination, the photons produced by these decays can reionize neutral hydrogen
in the intergalactic medium, with observable consequences for the CMB.~
Planck measurements of the CMB constrain the rate of electromagnetic energy deposited
into the CMB at or after the time of recombination $\trec \sim 10^{13}~\s$
(for reviews, see, \eg, Ref.~\cite{SlatyerReionization}).  For a particle $\chi$ with
a lifetime $\tau_\chi \gg \tnow$, the energy deposited in the CMB is
roughly equal to the energy $E_{\mathrm{inj}}$ injected when it decays.  In the case in
which this injected energy is transferred entirely to photons or charged particles,
Planck data imply a constraint $\Omega_\chi \Gamma_\chi \lesssim 3 \times 10^{-26}~\s^{-1}$
on the product of the abundance $\Omega_\chi$ and the decay width $\Gamma_\chi$ of $\chi$.
By contrast, when $\tau_\chi \sim \trec$, the energy deposited in the CMB may be significantly
lower in comparison to $E_{\mathrm{inj}}$ and the corresponding constraints on
$\Omega_\chi$ and $\Gamma_\chi$ are typically considerably weaker.

In the scenario considered in Ref.~\cite{SlatyerReionization}, essentially all of the
initial mass energy $m_\chi$ of each decaying $\chi$ particle is injected into photons
or charged particles at the moment when that particle decays.  Thus,
$E_{\mathrm{inj}}\approx m_\chi$.  By contrast, in our off-diagonal dark-matter scenario,
the energy injected into such particles through $\chi_2$ decay is $E_{\mathrm{inj}} \approx \Delta m$.
The corresponding bound on the lifetime of $\chi_2$ in our scenario is therefore
\begin{equation}
  \tau_2 ~\gtrsim~ \tau_\chi^{\mathrm{max}} \left(\frac{\Delta m}{m}\right)~,
\end{equation}
where $\tau_\chi^{\mathrm{max}}$ is the upper limit on $\tau_\chi$ taken from Fig.~11 of
Ref.~\cite{SlatyerReionization} for a decaying particle of mass $m$ and mass fraction
$f_2$.  
Since the band displayed in this figure represents the results of a survey over $m_\chi$ and over the decay channels $\chi\rightarrow\gamma\gamma$ and
$\chi\rightarrow e^+e^-$, we extract a value of $\tau_\chi^{\mathrm{max}}$ applicable to our scenario in the following way.  We note that the bound
on $\tau_\chi$ is weakest in the case in which $m_\chi$ --- and thus also $E_{\mathrm{inj}}$ --- is small and $\chi$ decays into photons rather
than $e^+e^-$ pairs.  In our off-diagonal dark-matter scenario, each $\chi_2$ likewise decays principally into photons, and the energy $E_{\mathrm{inj}}$
injected by each such decay is small.
For this reason, we use the value of $\tau_\chi^{\mathrm{max}}$
which correspond to the upper edge of this band in deriving our bound on $\tau_2$ for a given
mass fraction $f_2$.


\subsection{Co-annihilation Constraints}

The leading constraints on the annihilation (or co-annihilation) 
of dark-matter particles into photons are those derived from Fermi-LAT studies of the 
gamma-ray spectra Milky-Way dwarf spheroidals~\cite{FermiLATDwarfSearch}.  In 
single-component dark-matter models, this constraint implies an upper bound 
$\langle \sigma v \rangle_{\bar{\chi}\chi}^{\text{max}}(m_\chi)$ 
on the thermally averaged annihilation cross-section of the dark-matter particle 
$\chi$, which depends on its mass $m_\chi$.
       
Since the co-annihilation rates in Eq.~(\ref{eq:CoannXSecs}) are approximately 
independent of $\Delta m$ in the $\Delta m \ll m_1 \approx m_2$ regime, it is
straightforward to translate this bound into a bound on dark-matter co-annihilation
in our off-diagonal scenario.  Indeed, the only modification we must make is to
account for the fact that the initial state involves two different particles
with potentially different energy densities $\rho_1(\vec{x})$ and 
$\rho_2(\vec{x})$.  
In single-particle dark-matter models, the contribution to the photon flux
from any particular point in space is proportional to $\rhotot^2(\vec{x})$, 
where $\rhotot(\vec{x})$ is the total density of dark matter at that point.
By contrast, in our off-diagonal scenario, the corresponding contribution
is instead proportional to $\rho_1(\vec{x})\rho_2(\vec{x})$.
For simplicity, we shall assume that the halo profiles for 
$\chi_1$ and $\chi_2$ have the same shape --- \ie, that $\rho_1(\vec{x})$ and 
$\rho_2(\vec{x})$ depend on $\vec{x}$ in the same way and differ only in terms of 
overall normalization.  Moreover, we shall take the normalization 
factors for $\rho_1(\vec{x})$ and $\rho_2(\vec{x})$ to be proportional to the 
number-density fractions $f_1^{(T,V)}$ and $f_{2}^{(T,V)}$.  Under these 
assumptions, the Fermi-LAT constraint on our scenario takes the form      
\begin{equation}
  \langle \sigma v \rangle_{\bar{\chi}_1 \chi_2}^{(V,T)}(m) 
    ~<~ \frac{\langle \sigma v \rangle_{\bar{\chi}\chi}^{\text{max}}(m)}
    {2 f_{2}^{(T,V)} [1- f_{2}^{(T,V)}]}~.
  \label{eq:CoAnnLimitRaw}
\end{equation}
 
As discussed in Sect.~\ref{sec:DecayConstraints}, the lifetime of $\chi_2$ in the 
vector case is necessarily significantly longer than the current age of the universe.
In this case, the co-annihilation constraint on the parameter space of our scenario 
takes the form 
\begin{equation}
  \frac{12 m^2}{\pi \Lambda^4}  f_{2}^{(V)} \Big[1- f_2^{(V)}\Big] ~<~
    \langle \sigma v \rangle_{\bar{\chi}\chi}^{\text{max}}(m)~.
  \label{eq:CoAnnLimitVec}
\end{equation}
We note that this constraint depends on $\Lambda$, $m$, and $f_2^{(V)}$, but
not on the mass splitting $\Delta m$. 
By contrast, in the antisymmetric-tensor case, $\tau_2$ can be of 
order the age of the universe.  The co-annihilation constraint then takes the form
\begin{equation}  
  \frac{24 m^2}{\pi \Lambda^4} f_{2}^{(T)} (\tnow) 
    \Big[ 1- f_{2}^{(T)} (\tnow) \Big] ~<~
    \langle \sigma v \rangle_{\bar{\chi}\chi}^{\text{max}}(m)~.~~
  \label{eq:CoAnnLimitTens}
\end{equation}
In this case, since $f_{2}^{(T)}(\tnow)$ depends on $\Delta m$ through 
Eq.~(\ref{eq:f2Tens}), the co-annihilation constraint involves all four of our 
model parameters. 

Once again, we emphasize that the constraints in Eqs.~(\ref{eq:CoAnnLimitVec}) 
and~(\ref{eq:CoAnnLimitTens}) are derived under the assumption that the 
contact-operator description of the interaction between the dark and visible
sectors in 
Eqs.~(\ref{eq:LintVecTensRaw}) and (\ref{eq:LintVecTensMassEigs}) 
remains valid up to the energies
$\sqrt{s} \sim \mathcal{O}(m)$ relevant for co-annihilation.
However, we also note that Fermi-LAT data impose stringent constraints on 
dark-matter annihilation even in scenarios in which this is not the case --- for 
example, in scenarios involving a light $s$- and $t$-channel 
mediator~\cite{FermiBoundsWithMediators}.  Generally speaking, these bounds tend 
to be roughly similar to the bounds obtained in the contact-operator description.  
Nevertheless, care should be taken in interpreting the constraints in 
Eqs.~(\ref{eq:CoAnnLimitVec}) and~(\ref{eq:CoAnnLimitTens}) within the context 
of any particular UV theory.

\subsection{Direct-Detection Constraints\label{sec:DirectDetectionConstraints}}

Direct-detection experiments constrain the overall event rate $R$ in 
Eq.~(\ref{eq:rateGeneral}) for dark-matter scattering off atomic nuclei.
In principle, additional information about the scattering kinematics can 
also be extracted from the recoil-energy spectrum $dR/dE_R$ --- information that is
sensitive to the kinematic differences between elastic and inelastic scattering.
However, the detailed shape of the recoil-energy spectrum is also
sensitive to astrophysical properties of the dark-matter distribution in the 
Milky Way.  These include quantities such as the velocity distribution $\mathcal{F}(\vec{v})$, about 
which there are significant uncertainties.  We therefore restrict our attention
to bounding the total event rate, deriving an upper limit on $R$ in our off-diagonal 
dark-matter scenario from the applicable experimental results, and translating this
bound into a constraint on the allowed 
$\lbrace \Lambda, \Delta m, m, f_2^{(V,T)}\rbrace$
parameter space.

As discussed in Sect.~\ref{sec:DirectDetectionConstraints}, a vector interaction
in our off-diagonal dark-matter model contributes to SI scattering, while an 
antisymmetric-tensor interaction gives rise to SD scattering.  We consider
each of these cases in turn.  

The leading constraints on SI dark-matter scattering are those from the XENON1T, LUX, 
and PANDA-X experiments.
The most stringent limits from XENON1T are based on results obtained 
for $3.6 \times 10^4 ~\kg ~ \days$ of exposure~\cite{XENON1T}.  These results place 
an upper bound $R < 5 \times 10^{-5} \text{ events kg}^{-1}\,\text{day}^{-1}$ on 
the event rate for events within a recoil-energy range window 
$3 \, \kev \leq E_R \leq 50 \, \kev$ effectively determined by the detector-efficiency 
function $\epsilon(E_R)$ given in Ref.~\cite{XENON1T}.  The most stringent 
limits from LUX are 
based on results obtained with $3.4 \times 10^4 ~\kg ~ \days$ of exposure~\cite{LUXSI}.
For events within the recoil-energy window 
$1.1 \, \kev \leq E_R \lesssim 60 \, \kev$, where the upper end of this range 
is effectively determined by the efficiency function $\epsilon(E_R)$ given in 
Ref.~\cite{LUXSI}, the corresponding upper bound on the event rate is 
$R < 8 \times 10^{-5} \text{ events kg}^{-1}\,\text{day}^{-1}$.
In translating this limit into a constraint on the parameter space of our 
off-diagonal dark-matter scenario, we assume a Helm form factor~\cite{Helm,Engel}.
Recent results from PANDA-X~\cite{PANDAX} provide a slight improvement on the 
XENON1T limits on SI scattering for dark-matter particles of mass 
$m \gtrsim 100$~GeV.  However, this slight improvement does not have a significant 
impact on the region of our parameter space excluded by direct-detection experiments.  

In the case of SD scattering, the expected event rate at a given detector  
depends strongly on the relative strengths of the effective couplings of the
dark-sector particles to protons and neutrons.    
When the coupling to neutrons dominates, the most stringent limits
are once again those from the LUX experiment~\cite{LUXSD}, which provides a bound 
$R < 8 \times 10^{-5} \text{events kg}^{-1}\,\text{day}^{-1}$ on the event rate,
as discussed above.  By contrast, when the coupling to protons dominates, the most 
stringent limits are those from the PICO-60 experiment~\cite{PICO60}, which 
provides a bound $R < 1.7 \times 10^{-3} \text{ events kg}^{-1}\,\text{day}^{-1}$.  In
translating both the LUX and PICO-60 limits into bounds on our parameter space, we 
make use of the spin fractions and form factors from 
Ref.~\cite{KlosFormFactors}.  We take the efficiency function for PICO-60 from
Ref.~\cite{PICO60LEfficiency} and consider only the contribution to the
event rate from scattering events with recoil energies $E_R < 100$~keV at this 
detector.

In addition to assessing how current direct-detection limits constrain the  
parameter space of our scenario, we are also interested in determining the extent to
which future detectors might further probe that parameter space.  In particular,
we shall examine the reach provided by two hypothetical future detectors.  We model 
one of these detectors after the proposed LZ experiment~\cite{LZTDR}, which
provides increased sensitivity to both SI scattering and SD scattering 
in which the coupling to neutrons dominates.  We model the other detector after
the proposed PICO-500 experiment~\cite{PICO500}, which provides increased sensitivity 
to SD scattering in which the coupling to protons dominates. 
For the former detector, we assume an efficiency function $\epsilon(E_R)$ for the LZ 
detector which is identical to the efficiency function for LUX, but consider a 
slightly narrower recoil-energy window $6 \, \kev \leq E_R \leq 30 \, \kev$, in 
accord with in Ref.~\cite{LZTDR}.  This gives rise to a conservative estimate 
$R < 2 \times 10^{-6} \text{ events kg}^{-1}\,\text{day}^{-1}$ for the limit
on the scattering rate.  For the latter detector, we assume a sensitivity
equivalent to that of PICO-500~\cite{PICO500}, which
translates into a limit $R < 3 \times 10^{-5} \text{ events kg}^{-1}\,\text{day}^{-1}$ 
on the scattering rate, and an efficiency function identical to the efficiency function 
for PICO-60\cite{PICO60LEfficiency}.

\subsection{Collider Constraints\label{sec:ColliderConstraints}} 

The most relevant channels for dark-matter detection at hadron colliders in our
off-diagonal dark-matter scenario involve asymmetric pair-production through the process 
$q\bar{q} \rightarrow \chi_1\chi_2$ in association with one or more additional 
jets, with a photon, or with a $W^\pm$ or $Z$ boson.  Since we are assuming here that 
$\Delta m \lesssim \mathcal{O}(\mbox{MeV})$, the resulting phenomenology is
essentially indistinguishable from that associated with the analogous diagonal 
production processes in traditional single-component dark-matter models.  Constraints on these 
latter processes are therefore directly applicable to our scenario as well.  

The leading 
constraints on dark-matter production at hadron 
colliders are those from the analysis by the CMS 
collaboration~\cite{CMSMonoJetOrHadWZNew} at the $\sqrt{s}=13~\tev$ LHC with 
$35.9~\ifb$ of integrated luminosity, which combines results from both the  
${\rm monojet} + \slashed{E}_T$ and hadronically-decaying ${\rm W/Z} + \slashed{E}_T$ 
channels.  Searches in the ${\rm monojet} + \slashed{E}_T$~\cite{ATLASMonojet13TeVNeW} 
and hadronically-decaying ${\rm W/Z} + \slashed{E}_T$~\cite{ATLASMonoWZHadDec13TeV} 
channels at the same center-of-mass energy have also been performed 
by the ATLAS collaboration with $36.1~\ifb$ and $3.2~\ifb$ of integrated 
luminosity, respectively.
We note that searches for dark-matter production at colliders have been performed in the 
${\rm monophoton} + \slashed{E}_T$~\cite{ATLASMonoPhoton13TeV,CMSMonoPhoton13TeV} 
and leptonically-decaying mono-Z $+ \slashed{E}_T$~\cite{CMSMonoZLeptons13TeV} 
channels, as well as channels involving a single Higgs boson in conjunction 
with substantial $\slashed{E}_T$~\cite{ATLASMonoHiggsbb,CMSMonoHiggs,ATLASMonoHiggsGammaGamma}.
However, the constraints from these searches are generically subleading in comparison 
with the constraints from the monojet and hadronically-decaying mono-W/Z channels. 
Moreover, we note that since we are considering mass splittings 
$\Delta m \lesssim \mathcal{O}(\mbox{MeV})$, additional detection channels which 
can be relevant for off-diagonal dark-matter scenarios with larger mass splittings 
--- such as those involving a monojet and displaced pions $+ \slashed{E}_T$~\cite{BaiTait} 
or two energetic photons $+ \slashed{E}_T$~\cite{Yuhsin} --- 
do not constrain our parameter-space region of interest.     

In assessing the implications of these collider constraints on our off-diagonal 
dark-matter scenario, it is important to keep in mind that the contact-operator
description of the interactions between the dark and visible sectors is valid only
at energies comfortably below the scale $\Lambda$.  At higher energies, we would require
a more complete description of the full theory which at low energies gives rise to 
the effective operators in Eq.~(\ref{eq:LintVecTensRaw}).  For this
reason, a detailed analysis of the collider bounds on our scenario can only be
performed within the context of a particular such theory.  For example, the constraints quoted
in Ref.~\cite{CMSMonoJetOrHadWZNew} are derived for a set of simplified
models in which the dark and visible sectors interact via a massive scalar 
or vector mediator $\phi$ with a mass $m_\phi$.  

In this paper, by contrast, we shall seek to maintain generality by refraining from specifying a 
particular UV completion for the operators in Eq.~(\ref{eq:LintVecTensRaw}).
However, for reference, we shall nevertheless derive a heuristic bound on $\Lambda$ in 
both the vector and antisymmetric-tensor cases according to the following
procedure.  For a given choice of $m$ in our scenario, we compute the respective 
cross-sections $\sigma^{(V,T)}(pp \rightarrow \chi_1\chi_2j)$ and
$\sigma^{(V,T)}(pp \rightarrow \chi_1\chi_2 W/Z)$ for the processes which
contribute to the event rate in the ${\rm monojet} + \slashed{E}_T$ and hadronically-decaying 
${\rm W/Z} + \slashed{E}_T$ channels in our scenario after the imposition of 
the event-selection criteria outlined in Ref.~\cite{CMSMonoJetOrHadWZNew}.
In doing this we follow the procedure outlined in Sect.~\ref{sec:ColliderRates}. 
We then compare these cross-sections to the corresponding production cross-sections 
$\sigma^{(\phi)}(pp \rightarrow \chi\chi j)$ and
$\sigma^{(\phi)}(pp \rightarrow \chi\chi W/Z)$ obtained for the vector-mediator
model considered in Ref.~\cite{CMSMonoJetOrHadWZNew} after the imposition of
the same cuts.  Since $\sigma^{(V,T)}(pp \rightarrow \chi\chi j)\propto \Lambda^4$,
a lower limit $m_\phi \gtrsim m_{\phi,\mathrm{max}}$ on the mediator 
mass from the ${\rm monojet} + \slashed{E}_T$ channel corresponds to a bound
\begin{eqnarray}
  \Lambda &\gtrsim& 2 m_{\phi,\mathrm{max}}
    \left[\frac{\sigma^{(V,T)}(pp\rightarrow \chi_1\chi_2 j)}
    {\sigma^{(\phi)}(pp\rightarrow \chi\chi j)} \right]^{1/4}
  \label{eq:BoundOnLambda}
\end{eqnarray}
on the scale $\Lambda$.  
Note that the factor of two appearing in this expression
reflects the difference between our benchmark values for the operator 
coefficients $c_q^{(V,T)}$ and the benchmark values adopted for couplings in the 
CMS analysis.  A completely analogous bound on $\Lambda$ can likewise
be obtained from the hadronically-decaying ${\rm W/Z} + \slashed{E}_T$ channel.

Once again, we emphasize that such bounds on $\Lambda$ 
strictly apply only in the regime in which the contact-operator description of the 
interactions between the dark and visible sectors remains valid up to the energy scales 
$\sqrt{s}\sim  \mathcal{O}(\mbox{TeV})$ relevant for LHC physics.  
By contrast, if this assumption does not hold --- for example, if the contact interactions
in Eq.~(\ref{eq:LintVecTensRaw}) arise due to a light mediator particle which could be 
produced on shell at the LHC --- the collider constraints become highly model-dependent. 
Indeed, in many cases they become significantly weaker.  Moroever, we note that
there exists a systematic uncertainty of roughly $40\%$ in our signal-event rates, owing to 
hadronic physics and soft-QCD effects.  However, we also note that the corresponding 
uncertainty in our bound on $\Lambda$ is only around $10\%$.

Finally, we note that for larger values of $\Delta m$ than those considered here, a variety of additional collider signatures can arise.
For example, in this paper we have 
focused on the scenario in which the lifetime of $\chi_2$,
though potentially shorter than the age of the Universe, is nevertheless
long enough that it will not decay within the detection volume of the LHC.~ 
For larger values of $\Delta m$, however, 
the decay of the heavier component is more rapid and can thus occur
 {\it within}\/ the detection volume.
Such scenarios  have been considered in the context of
colliders~\cite{Bainew, Weinernew, Izagnew, Codex,Codex2,Codex3, Buchnew} as well as
fixed-target experiments~\cite{Morrisseynew, Izagnew2}.   Key features of these
scenarios are the detection of Standard-Model particles arising from
the decay of the heavier dark particle, potentially with a displaced vertex.


\section{Results: ~A Picture of Off-Diagonal Complementarity\label{sec:Results}}


\begin{figure}
\centering
    \includegraphics[width=0.44\textwidth]{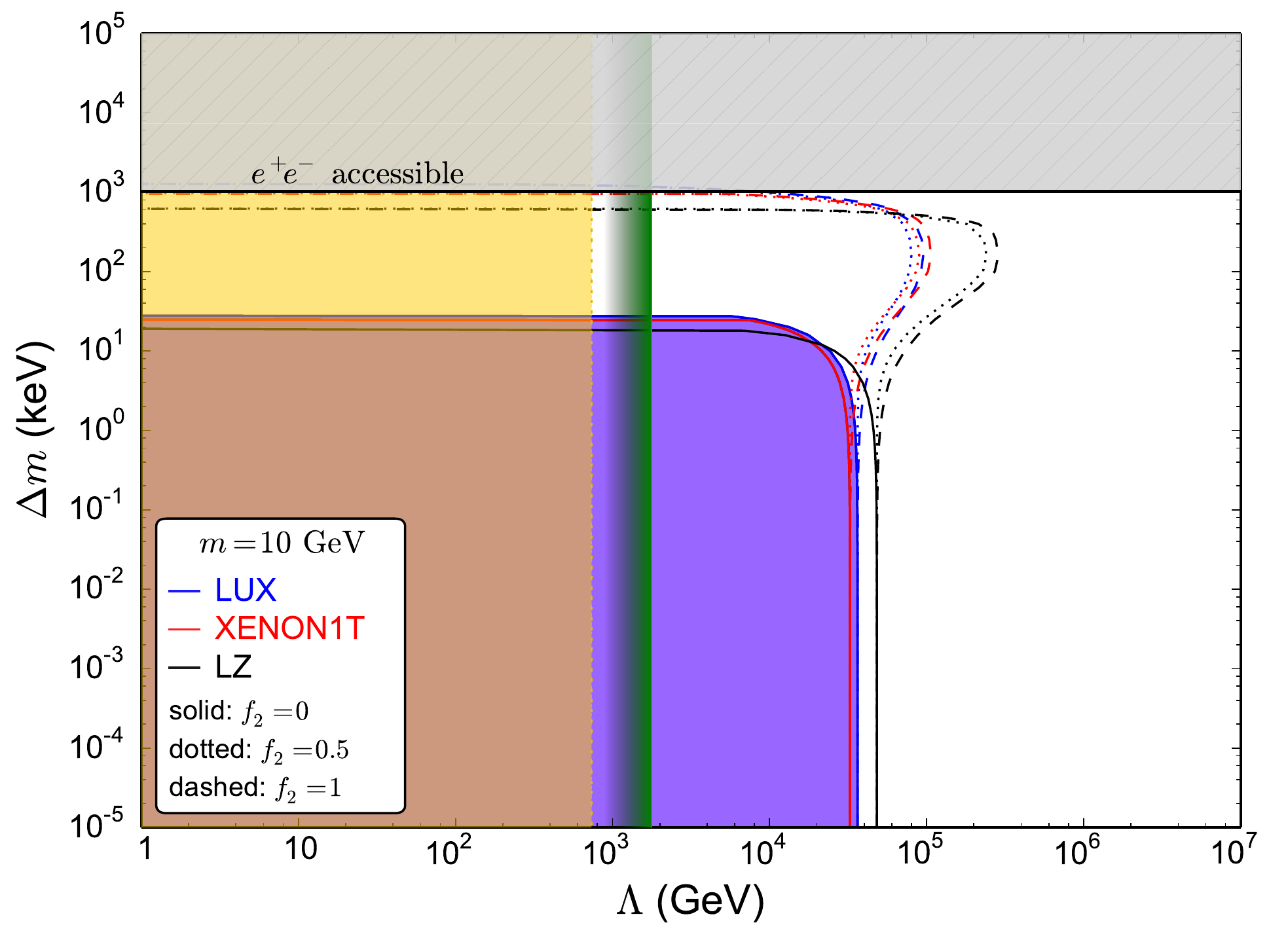}
    \includegraphics[width=0.44\textwidth]{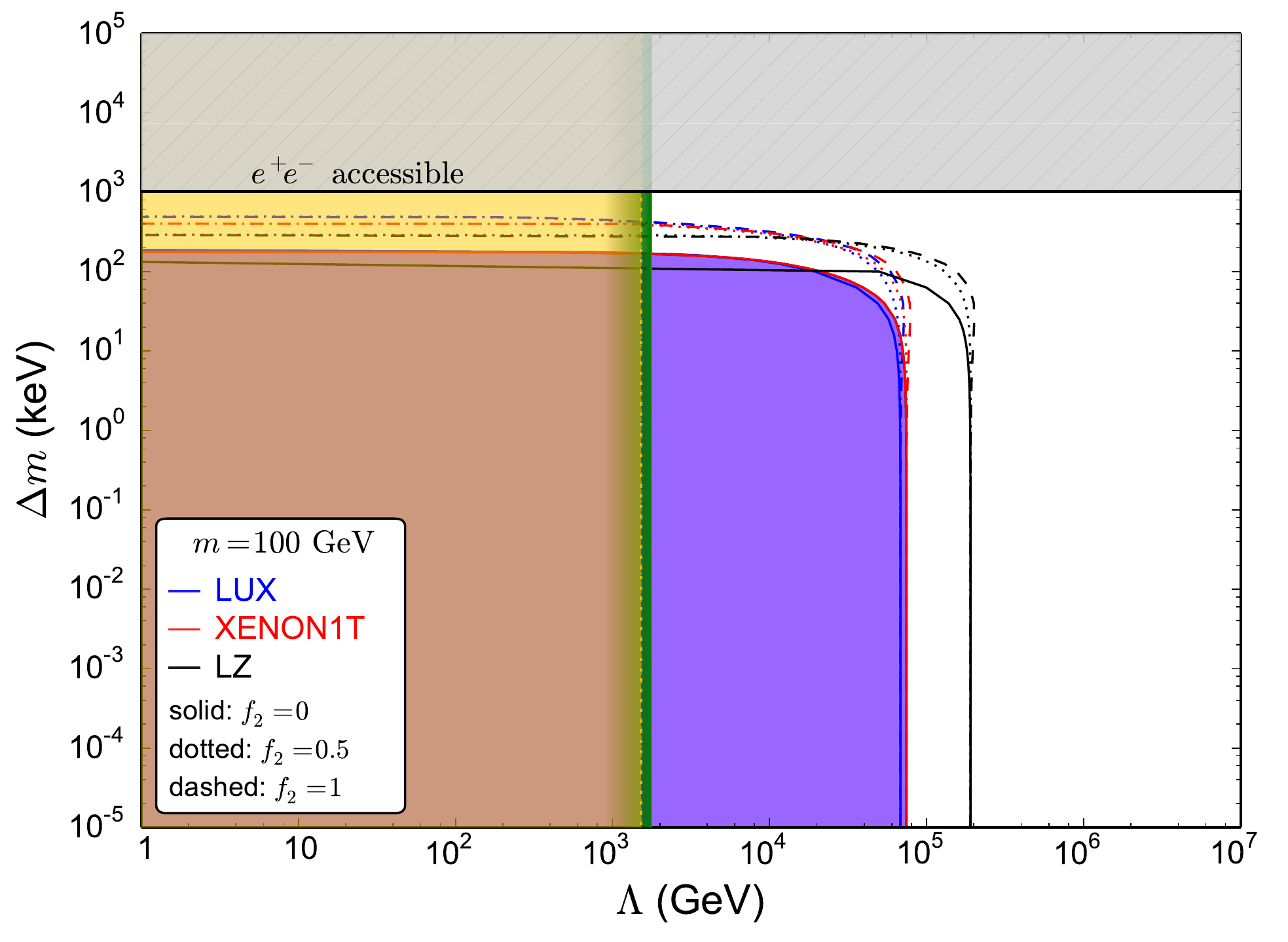}
    \includegraphics[width=0.44\textwidth]{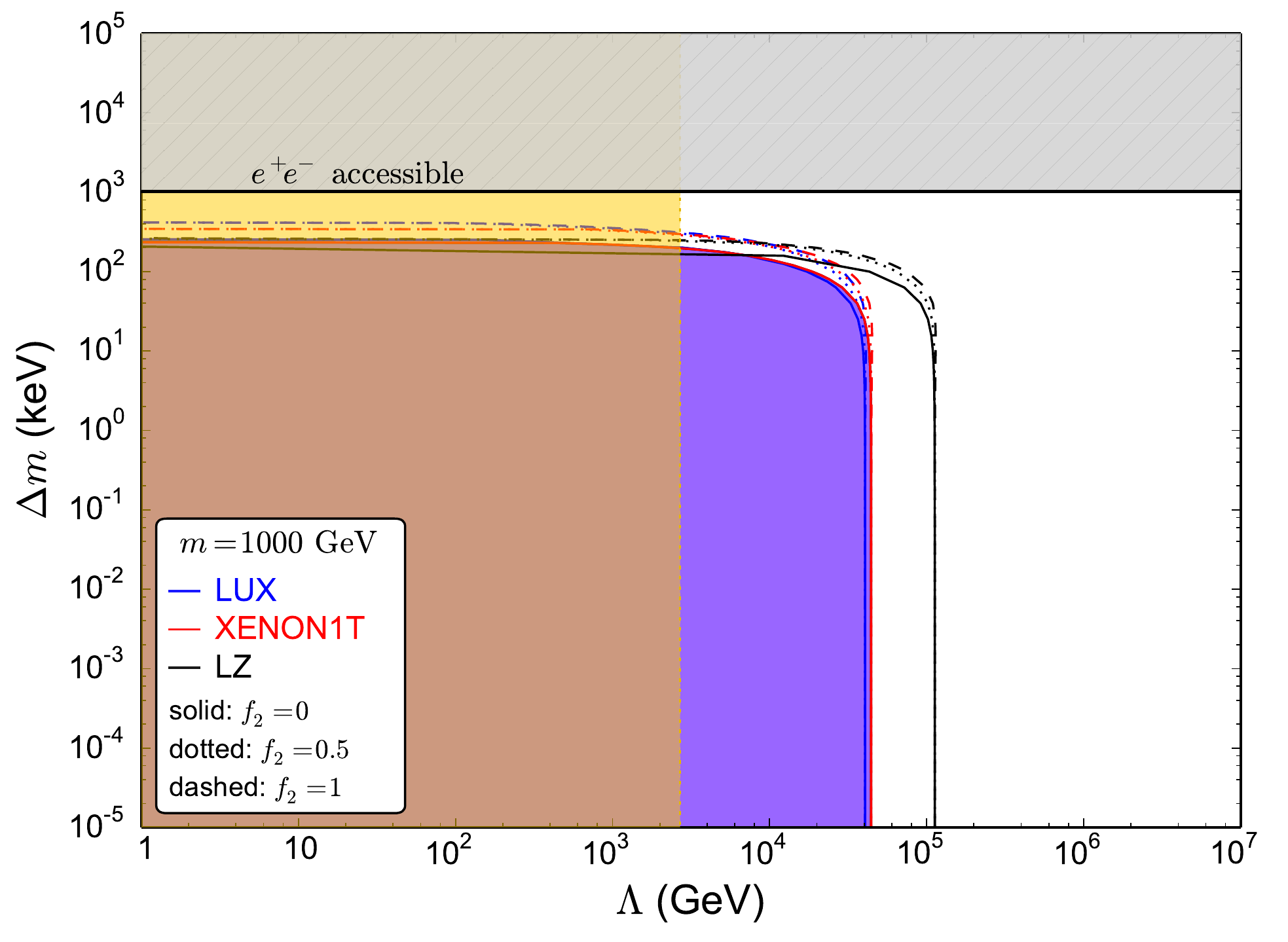}
\caption{Constraints on the parameter space of our scenario for the case of 
  the vector interaction in Eq.~(\protect\ref{eq:LintVecTensMassEigs}).  The 
  upper, middle, and lower panels correspond respectively to the choices
  $m = \{ 10, 100, 1000\}~\gev$.  The yellow shaded region is excluded by 
  co-annihilation limits from Fermi-LAT for $f_2^{(V)}=0.5$.
  The magenta shaded region is excluded by combined direct-detection limits from
  LUX and XENON1T for the conservative case of $f_2^{(V)}=0$.  Individual 
  constraint contours from both LUX and XENON1T, as well as contours representing the 
  projected reach of LZ, are shown for $f_2^{(V)}=0$ (solid curves), $f_2^{(V)}=0.5$ 
  (dotted curves), and $f_2^{(V)}=1$ (dashed curves).  The green line represents 
  the na\"{i}ve bound from CMS searches in contact-operator approximation. 
  For further details, see text.
 \label{fig:delmvsLambdaVECT}}
\end{figure}

\begin{figure}
\centering
    \includegraphics[width=0.44\textwidth]{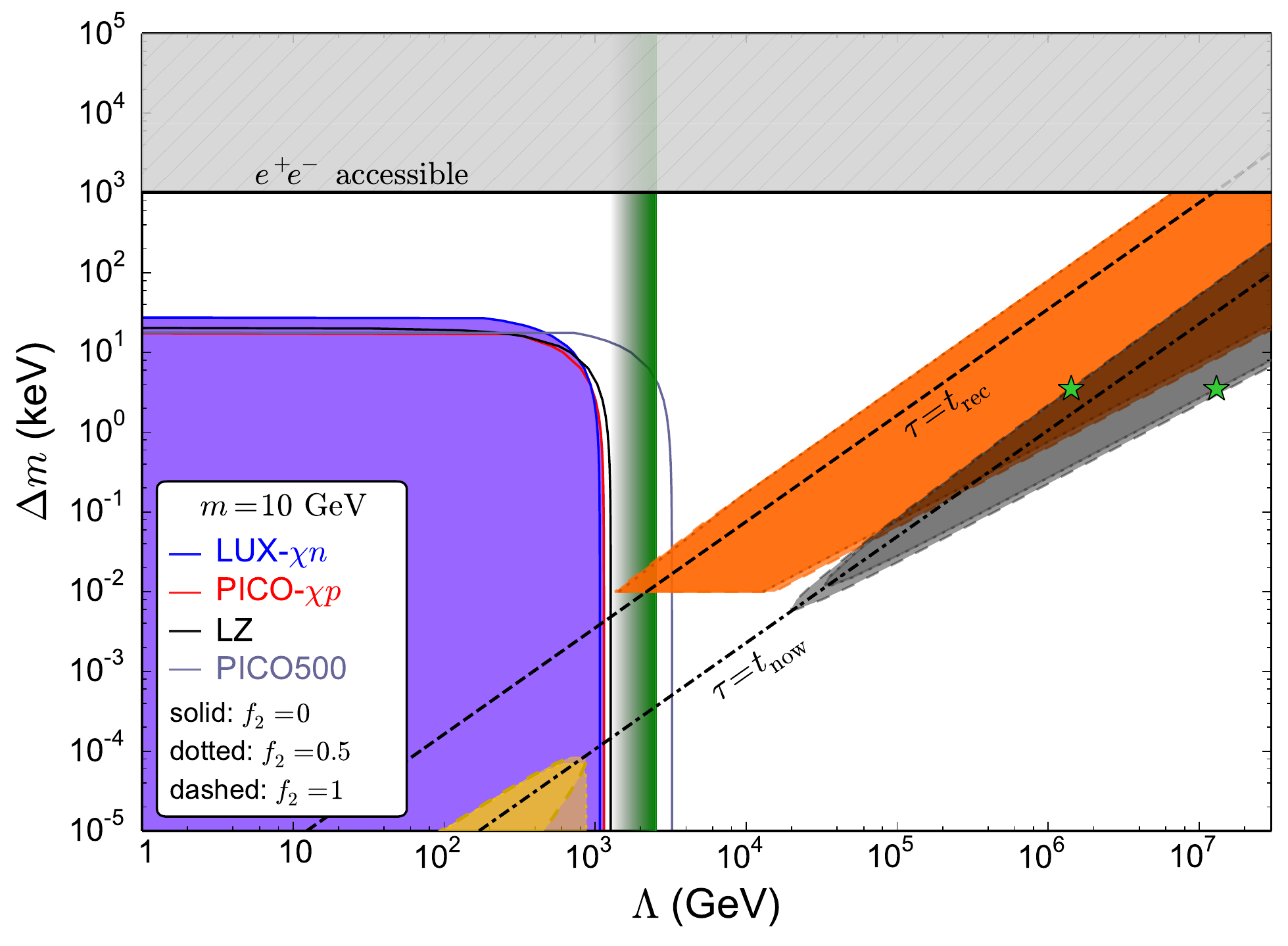}
    \includegraphics[width=0.44\textwidth]{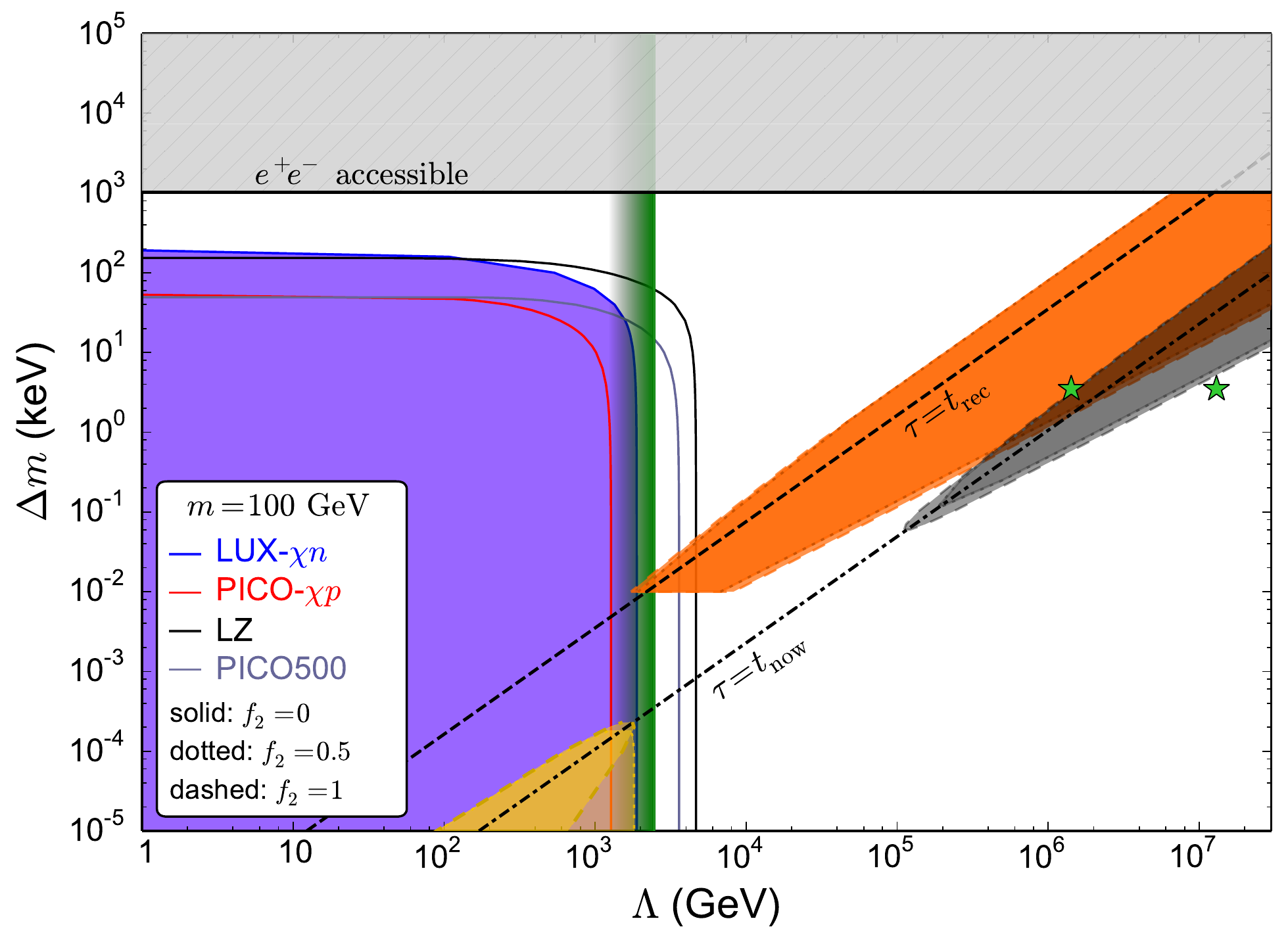}
    \includegraphics[width=0.44\textwidth]{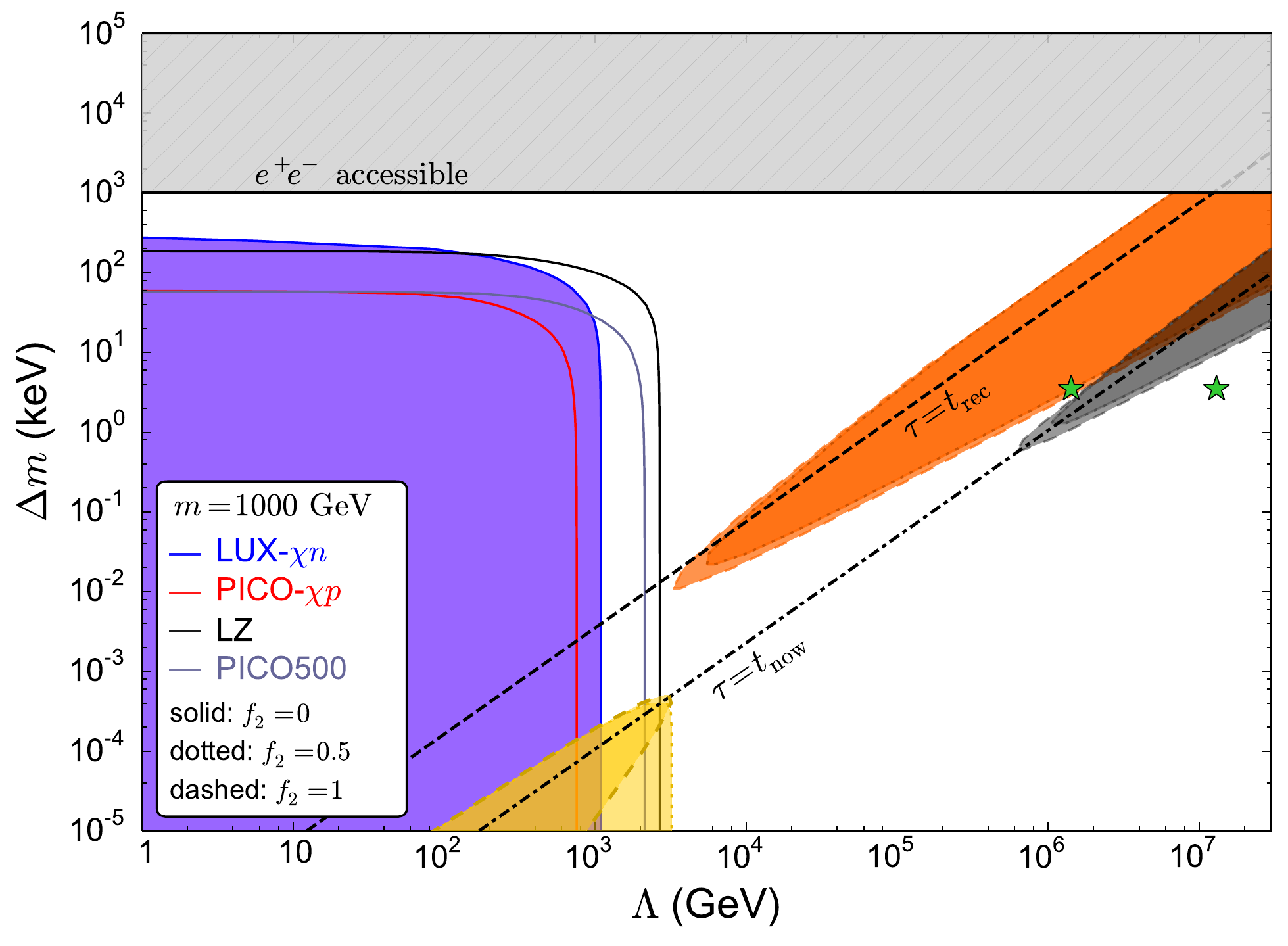}
\caption{Constraints on the parameter space of our scenario for the case of 
  the antisymmetric-tensor interaction in Eq.~(\protect\ref{eq:LintVecTensMassEigs}).
  The upper, middle, and lower panels correspond respectively to the choices
  $m = \{ 10, 100, 1000\}~\gev$.  The yellow shaded regions are excluded by 
  co-annihilation limits from Fermi-LAT, while the orange shaded regions
  are excluded by reionization limits and the gray shaded regions
  are excluded by X-ray searches. 
  The magenta shaded region is excluded by combined direct-detection limits from
  LUX and PICO-60.  Individual constraint contours from both LUX and PICO-60, as 
  well as contours representing the projected reach of both LZ and PICO-500, are 
  also shown.  Note that the direct-detection curves for all benchmark
  choices of $f_2^{(T)}$ effectively coincide.  The green line represents 
  the na\"{i}ve bound from CMS searches in contact-operator approximation. 
  The green stars indicate the points in parameter 
  space consistent with the purported $3.5 \, \kev$ line observed in 
  galactic clusters and in the Milky Way and Andromeda halos.  For further details, 
  see text.  
\label{fig:delmvsLambdaTENS}}
\end{figure}

We now turn to examine how the experimental limits discussed in Sect.~\ref{sec:Constraints}
collectively constrain the parameter space of our off-diagonal dark-matter scenario.
We present our results as constraint contours in the $(\Lambda,\Delta m)$ plane for a
set of benchmark values $m = \{10\, \gev, 100\, \gev, 1\, \tev \}$ for the dark-matter
mass scale.  A set of different contours corresponding to different choices 
$f_2^{(V,T)}=\{0,0.5,1\}$ for the primordial abundance fraction is included for the
decay, co-annihilaion, and direct-detection constraints, which are sensitive to the 
value of $f_2^{(T,V)}$.  
The results for the case of a vector interaction are shown in 
Fig.~\ref{fig:delmvsLambdaVECT}, while the results for the case of an antisymmetric-tensor 
interaction are shown in Fig.~\ref{fig:delmvsLambdaTENS}.  

The yellow shaded regions correspond to the exclusion regions associated 
with the Fermi-LAT \verb+PASS8+ co-annihilation bounds for $f_2^{(V,T)} = 0.5$
(light yellow) and $f_2^{(V,T)} = 1$ (dark yellow).  The orange shaded regions 
correspond to the exclusion regions from reionization effects on the CMB for 
$f_2^{(V,T)} = 0.5$ (dark orange) and $f_2^{(V,T)} = 1$ 
(light orange).  The gray shaded regions correspond to the exclusion 
contours from X-ray line searches for $f_2^{(V,T)} = 0.5$ (dark gray) and 
$f_2^{(V,T)} = 1$ (light gray).  
Constraint contours corresponding to bounds from individual direct-detection 
experiments are shown for $f_2^{(V,T)}=0$ (solid lines), $f_2^{(V,T)}=0.5$ 
(dotted lines), and $f_2^{(V,T)}=1$ (dashed lines).  Projected limits for future
direct-detection experiments are displayed in the same way.  The purple shaded region 
corresponds to the region which is excluded by  the most conservative case, wherein 
$f_2^{(V,T)}=0$ and the signal contribution is due to up-scattering alone. 
The green line appearing in the $m = 10$~GeV and $m=100$~GeV panels of 
Fig.~\ref{fig:delmvsLambdaVECT} and~\ref{fig:delmvsLambdaTENS} corresponds
to the combined upper limit derived from ${\rm monojet} + \slashed{E}_T$ and 
hadronically-decaying ${\rm W/Z} + \slashed{E}_T$ searches at CMS for the 
case of a contact interaction.  This line is omitted in the $m = 1$~TeV
panels, since the contact-operator description is not valid within the region
of parameter space constrained by these searches for such a large value of $m$.

The striped gray region at the top of each panel of 
Figs.~\ref{fig:delmvsLambdaVECT} and~\ref{fig:delmvsLambdaTENS} 
indicates the region of parameter 
space within which decay channels for $\chi_2$ involving $e+e^-$ pairs in the final 
state are kinematically accessible.  We emphasize that while this region lies 
outside our regime of interest, it is not necessarily excluded in its entirety. 
Finally, the green stars in Fig.~\ref{fig:delmvsLambdaTENS} indicate the points in 
parameter space at which our scenario is capable of explaining the $3.5$-keV X-ray line.

As discussed near the beginning of
Sect.~\ref{sec:Constraints},
the parameter $\Delta m$ captures 
(and in some sense quantifies) the off-diagonality  
inherent in our dark-matter scenario. 
Indeed, as $\Delta m\to 0$, we recover the bounds and constraints that would be 
expected for a ``diagonal'', single-component dark-matter scenario.
Our main interest, therefore,  is in determining how the behavior of these constraints 
evolves --- and what new constraints may appear --- as $\Delta m$ increases from zero
within each of the panels in
Figs.~\ref{fig:delmvsLambdaVECT} and~\ref{fig:delmvsLambdaTENS}.

\subsection{Vector interaction}

The constraint contours for the case of a vector interaction are shown in 
Fig.~\ref{fig:delmvsLambdaVECT}.  Since $\chi_2$ is extremely stable in 
this case, as discussed in Sect.~\ref{sec:DecayConstraints}, experimental limits
from dark-matter decay --- including both those from X-ray instruments and those from
reionization effects on the CMB --- yield no meaningful bounds on our parameter space.

The longevity of $\chi_2$ in the vector case also has an important impact on
co-annihilation constraints.  Co-annihilation requires that a non-negligible 
population of both $\chi_1$ and $\chi_2$ be present in the dark-matter halo of the 
astrophysical object under observation.  Indeed, Eq.~(\ref{eq:CoAnnLimitRaw}) implies 
that the co-annihilation constraint on the parameter space of our scenario is the most 
severe when equal numbers of $\chi_1$ and $\chi_2$ are present in the halo and 
disappears altogether when only one species is present.  Since $\chi_2$ is 
stable on cosmological timescales in the vector case, we have 
$f_2^{(T,V)}(\tnow) \approx f_2^{(T,V)}$. As a result, the co-annihilation 
constraint contours for both $f_2^{(T,V)}=0$ and $f_2^{(T,V)}=1$ vanish.  Moreover,
since the constraint is essentially independent of $\Delta m$ in this case, the  
constraint contour for $f_2^{(T,V)} = 0.5$ appearing in each panel of 
Fig.~\ref{fig:delmvsLambdaVECT} is effectively a vertical line.

For small $\Delta m$, we find that direct-detection constraints
generically provide the strongest bounds on $\Lambda$. 
However, we see that direct-detection experiments quickly lose sensitivity once 
$\Delta m$ exceeds a certain threshold.  This behavior is the result of a non-trivial interplay 
between scattering kinematics and detector-performance considerations. 
For up-scattering, the additional energy $\Delta m$ required to produce the final-state 
$\chi_2$ particle must be supplied by the kinetic energy of the incoming $\chi_1$ particle.
Since the velocity distribution $\mathcal{F}(\vec{v})$ for dark-matter in the Milky-Way 
halo is suppressed at large $v$ --- and indeed drops to zero for 
$v \geq v_{\mathrm{esc}}$ --- this energy threshold becomes prohibitively large for 
sufficiently large $\Delta m$.  By contrast, for down-scattering, the scattering process 
is exothermic, with the additional mass energy $\Delta m$ released by the incoming 
$\chi_2$ providing an additional contribution to the kinetic energies of the final-state 
particles.  For this reason, we find that the presence of even a tiny present-day 
number-density fraction $f_2^{(V)}(\tnow) \approx f_2^{(V)}$ for the heavier dark-matter 
species results in a dramatic increase in sensitivity in the region above the $\Delta m$ 
threshold for up-scattering relative to the sensitivity obtained for $f_2^{(V)}(\tnow)=0$.

Indeed, it is apparent from Fig.~\ref{fig:delmvsLambdaVECT} that direct-detection 
constraints are typically strongest when 
down-scattering dominates and $\Delta m \sim \mathcal{O}(1 - 100~\kev)$, as is 
apparent in the constraint contours obtained for $f_2^{(T,V)} = 1$. 
However, for sufficiently large $\Delta m$, typical $E_R$ 
values associated with down-scattering events lie above the recoil-energy window associated 
with these experiments --- a window whose upper limit is effectively determined by efficiency
considerations and event-selection requirements.  Thus, for down-scattering as well
as up-scattering, there exists an upper limit on the value of $\Delta m$ to which 
direct-detection experiments are sensitive.   

As discussed in Sect.~\ref{sec:ColliderConstraints}, collider constraints on
our scenario are essentially independent of $\Delta m$ for $\Delta m \ll m$.
Consequently, the corresponding constraint contours in the panels of 
Fig.~\ref{fig:delmvsLambdaVECT} also appear as vertical lines.  For $m=10$~GeV, 
the constraints are roughly comparable to those from co-annihilation.  However,
as $m$ increases, the collider constraints become weaker while the co-annihilation
constraints become stronger.  
The decrease in collider sensitivity is to be expected, given 
that the pair-production rate at colliders decreases with increasing $m$ when
$m$ is large.  On the other hand, the increase in indirect-detection sensitivity to 
co-annihilation is due primarily to an increase in the energy of the visible particles 
produced by co-annihilation.    
We note that while the number densities of our dark-matter species decrease 
with increasing $m$, this suppression is compensated by an
increase in the co-annihilation cross-section.

In summary, for the case of a vector interaction, direct-detection constraints 
provide the most stringent bounds on --- and the best projected reach 
within --- the parameter space of our scenario.  For $\Delta m \lesssim 100$~keV,
current limits from LUX and XENON1T exclude values of $\Lambda$ up to 
$10$ - $100$~TeV.~  However, we also see that for larger mass splittings in
the range $100~\kev \lesssim \Delta m \lesssim 1~\mev$, indirect-detection 
bounds from co-annihilation play an important role, filling in the ``gap'' between
where direct-detection experiments lose sensitivity --- especially when 
$f_2^{(V)}$ is small --- and where dark-matter decay channels to final 
states involving $e^+e^-$ pairs open up.     

\subsection{Tensor interaction}

The constraint contours for the case of a tensor interaction are shown in 
Fig.~\ref{fig:delmvsLambdaTENS}.  We observe that in this case far shorter lifetimes 
for $\chi_2$ can be realized within our parameter-space region of interest than in the 
case of a vector interaction.  The behavior of $\tau_2$ within the $(\Lambda,\Delta m)$
plane can be inferred from Eq.~(\ref{eq:tau2Tens}), which indicates that    
$\tau_2 \propto \Lambda^4 / (\Delta m)^3$.  Thus, we see that lifetime of $\chi_2$
increases as one moves downward and to the right in each panel of 
Fig.~\ref{fig:delmvsLambdaTENS}, and that contours of constant $\tau_2$ are 
straight lines running from the lower left to the upper right.  For reference, we 
explicitly include two such contours (the thick dashed lines) on each panel of
this figure.  The first corresponds to a lifetime $\tau_2 = \tnow$, which is the 
timescale relevant for X-ray line searches.  The second corresponds to a lifetime 
$\tau_2 = \trec$, which is the timescale relevant for CMB physics.

We see that the region of our parameter space most strongly constrained by 
X-ray line searches in each panel of Fig.~\ref{fig:delmvsLambdaTENS} is that in which 
$\tau_2 \sim \tnow$.  The corresponding constraint contours, which are given by
Eq.~(\ref{eq:GamTensBound}), depend on the fundamental scales $\Lambda$, $\Delta m$, and $m$
which characterize our scenario through $\tau_2$ and through the ratio 
$\Delta m/m$, which represents the fraction of the mass energy of $\chi_2$ which is 
transferred to the kinetic energies of the decay products.  Indeed, within each 
panel --- \ie, for fixed $m$ --- we see that the X-ray constraints grow weaker as 
$\Delta m$ decreases.  Likewise, comparing the different panels of the figure, we also 
see that these constraints grow weaker as $m$ increases.  Moreover, we 
also see that the X-ray constraints become weaker as $f_2^{(T)}$, which specifies 
the number density of $\chi_2$ in the early universe, decreases.         

By contrast, we see that the region of parameter space most strongly constrained by
reionization limits in each panel of Fig.~\ref{fig:delmvsLambdaTENS} is that in
which $\tau_2$ is near or slightly above $\trec$.
For $\tau_2 \lesssim \trec$, the majority of $\chi_2$ particles decay well before
recombination; thus, the energy injected by those decays does not contribute to
reionization effects on the CMB.~  By contrast, for $\tau_2 \gtrsim \trec$, the majority
of $\chi_2$ particles decay after recombination and stringent constraints on our
parameter space arise.  Indeed, for lifetimes in the range
$\trec \lesssim \tau_2 \lesssim \tnow$, these constraints are competitive with the
X-ray constraints discussed above, whereas for $\tau_2 \gg \tnow$, those X-ray
constraints dominate.  Once again, we observe that the overall shape of the constraint
contours from reionization is essentially determined by the interplay between $\tau_2$ and
the ratio $\Delta m/m$.  As a result, this shape is similar to the overall shape of
the X-ray contours.  The only salient difference is that the reionization contours
are cut off below $\Delta m \sim 13.6$~eV, since a photon with $E_\gamma$ below this
threshold cannot ionize neutral hydrogen.

We observe that the region of our parameter space in Fig.~\ref{fig:delmvsLambdaTENS} 
excluded by Fermi-LAT limits on dark-matter co-annihilation differs significantly
from the region excluded in Fig.~\ref{fig:delmvsLambdaVECT}.  This is ultimately
due to the fact that co-annihilation requires a significant population of $\chi_2$ to 
be present in the halos of such galaxies at present time. 
Thus, just as with the X-ray constraints discussed above, we find that 
co-annihilation limits place no meaningful bounds on our parameter space when 
$\tau_2 \ll \tnow$.  However, for $\tau_2 \gtrsim \tnow$, an observable 
co-annihilation signal may indeed arise.   
In situations in which a substantial primordial abundance is generated for 
both $\chi_1$ and $\chi_2$ --- as is the case, for example, for our
number-density fraction benchmark $f_2^{(T)} = 0.5$ --- the co-annihilation 
bound on $\Lambda$ from Eq.~(\ref{eq:CoAnnLimitTens}) is not particularly 
sensitive to the value of $\Delta m$.

By contrast, in situations in which the primordial abundance of $\chi_1$ is
negligible and essentially all of the dark matter is initially in the $\chi_2$
state --- as is the case for our number-density fraction benchmark 
$f_2^{(T)} = 1$ --- a substantial co-annihilation rate can only be achieved
if a non-negligible population of $\chi_1$ particles is subsequently generated
by $\chi_2$ decays on cosmological time scales.  On the one hand, this means that 
if $\tau_2 \gg \tnow$, essentially all of the dark matter will still be in the 
$\chi_2$ state at present time, and the co-annihilation rate will be negligible.
On the other hand, if $\tau_2 \ll \tnow$, essentially all of the dark matter will
be in the $\chi_1$ state at present time, as discussed above, and the co-annihilation 
rate will likewise be negligible.  Indeed, for $f_2^{(T)} \approx 1$, a lifetime 
$\tau_2 \sim \tnow$ is required in order to achieve a non-negligible 
co-annihilation signal.  In this case, the corresponding constraint contours
exhibit a clear sensitivity to the value of $\Delta m$, as evident in 
Fig.~\ref{fig:delmvsLambdaTENS}.    

As discussed in Sect.~\ref{sec:DirectDetection}, the primary difference between 
the direct-detection phenomenology which stems from a vector interaction and
that which stems from an antisymmetric-tensor interaction is that the former
gives rise to SI scattering, whereas the latter only gives rise to SD scattering.
Since the limits on SI scattering are considerably more stringent than those
on SD scattering, the direct-detection constraints on $\Lambda$ in 
Fig.~\ref{fig:delmvsLambdaTENS} are significantly weaker than those in 
Fig.~\ref{fig:delmvsLambdaVECT}.  However, another important difference arises due
to a difference between the characteristic ranges $\tau_2$ associated 
with the vector and tensor interactions in our off-diagonal dark-matter scenario. 
Throughout almost all of the parameter space in Fig.~\ref{fig:delmvsLambdaTENS} 
within which direct-detection experiments are sensitive to inelastic scattering
in our scenario, we have $\tau_2 \ll \tnow$.  This implies that essentially all of the
the dark-matter in the Milky-Way halo at present time is in the $\chi_1$ state.
The only non-negligible contribution to the event rate at direct-detection 
experiments is therefore due to up-scattering, which is kinematically 
suppressed, as discussed above --- especially for large $\Delta m$.
The only region of parameter space in which $\chi_2$ is cosmologically stable
and in which one might expect a direct-detection signal is a region in which $\Delta m$
is so small that there is essentially no kinematic difference between up-scattering 
and down-scattering.  Thus, the direct-detection phenomenology that arises in the
case of an antisymmetric-tensor interaction is essentially independent of
$f_2^{(T)}$, and the constraint contours for all three of our benchmark choices
for this parameter coincide.

The collider constraints on our parameter space 
in the antisymmetric-tensor case are independent of $\Delta m$, as they were in the 
vector case.  Thus, the corresponding constraint contours in the panels of 
Fig.~\ref{fig:delmvsLambdaTENS} likewise appear as vertical lines.  However, we
see that in the case of an antisymmetric-tensor interaction, collider constraints and 
the direct-detection constraints are quite competitive.  Indeed, we see that for 
$\Delta m \lesssim 10$~keV, the combined monojet and hadronically-decaying W/Z bounds
from ATLAS currently represent the leading constraint on $\Lambda$.  However, we also
see that the projected reach of the LZ experiment will supersede this.  Perhaps
even more importantly, however, colliders represent essentially the only probe of 
the region of parameter space in which $10~\kev \lesssim \Delta m \lesssim 1~\mev$
and $\Lambda \lesssim 1$~TeV.~  

In summary, in the case of a tensor interaction, a picture of dark-matter complementarity 
emerges in which experimental probes of the dark sector mostly cover different,
non-overlapping regions of our parameter space.  For $\Delta m \lesssim 100$~keV and 
$\Lambda \lesssim 1$~TeV, direct-detection experiments provide excellent coverage 
of this parameter space, and results from LZ and PICO-500 are projected to extend
that coverage over significantly higher values of $\Lambda$ in the near future.
For larger values of $\Lambda$, indirect detection plays the dominant role in probing
our parameter space.  Gamma-ray detectors sensitive to dark-matter 
co-annihilation provide some coverage for smaller $\Lambda$ and $\Delta m$, while probes 
of dark-matter decay, including both X-ray line searches and observations of the CMB, 
provide coverage for larger values of these paramaters.  

One salient difference between
the complementarity picture which emerges in the antisymmetric-tensor case, in comparison with the one
which emerges in the vector case, is that there exists a range of mass splittings 
$100~\kev \lesssim \Delta m \lesssim 1~\mev$ for which the only bounds on $\Lambda$
are those from colliders.  However, as discussed in Sect.~\ref{sec:ColliderConstraints}, 
these bounds are highly model-dependent.  These considerations motivate efforts to explore 
this region of parameter space using other, complementary probes of the dark sector as 
well.  One promising possibility is that by extending the range of nuclear-recoil energies 
accessible at direct-detection experiments to higher $E_R$, future such experiments could 
provide more robust coverage of some or even all of this region.


\section{Conclusions and Outlook \label{sec:Conclusions}}


Recent years have seen increasing focus on multi-component dark sectors.
Indeed,
a particularly dramatic example of this is the so-called Dynamical Dark Matter framework 
of Refs.~\cite{DDM1,DDM2,DDM3}.
However, such multi-component dark sectors give rise
to two general classes of dark-matter processes:  ``diagonal'' processes that involve only 
dark-matter species, and ``off-diagonal'' processes that involve two (or more)
dark-matter species.
Normally, one might expect the experimental signals from 
diagonal processes to dominate those from off-diagonal processes.
However, as we have seen, there exist particular multi-component dark-matter
scenarios in which the diagonal processes are absent or suppressed, and in which
it is the {\it off}\/-diagonal processes which dominate the resulting phenomenology.
 
In this paper, we have examined the phenomenology of such off-diagonal dark-matter 
scenarios --- scenarios in which the dark sector couples to the visible sector
primarily through interactions involving two different dark-sector fields.
Scenarios of this sort give rise to inelastic scattering at direct-detection
experiments, dark-matter co-annihilation in the halos of galaxies, asymmetric 
pair-production processes at colliders, and dark-matter decay.
We have shown that such off-diagonal dark-matter scenarios naturally arise
when the dark sector
consists of a Dirac fermion $\chi$ which is subsequently split into a pair of 
nearly degenerate mass eigenstates $\chi_1$ and $\chi_2$ by a small Majorana mass.    
In such ``pseudo-Dirac'' models, if $\chi$ couples to the visible sector primarily through operators
involving the vector bilinear $\bar{\chi}\gamma^\mu\chi$ or the tensor bilinear
$\bar{\chi}\sigma^{\mu\nu}\chi$, the resulting couplings between these mass 
eigenstates and the visible sector are purely off-diagonal.  We have examined 
the phenomenology which follows from such scenarios both in the case of a vector 
interaction and in the case of an antisymmetric-tensor interaction, and we     
have examined the picture of dark-matter complementarity which arises in each case.
 
A few comments are in order.  First, we emphasize that for sake of generality 
we have refrained from specifying a UV completion for either of the contact operators 
appearing in Eq.~(\ref{eq:LintVecTensMassEigs}).
For this reason, the collider constraints we have taken into account in constraining 
our off-diagonal dark-matter scenario are solely those associated with model-independent 
dark-matter detection channels such as ${\rm monojet} + \slashed{E}_T$, 
${\rm monophoton} + \slashed{E}_T$, \etc, which pertain to the contact-operator
description.  In the regime in which $\Lambda \lesssim \mathcal{O}(\tev)$, the 
contact-operator formulation is no longer reliable and other channels may 
play an important role in the collider phenomenology of the scenario.  For example,
if the relevant contact operator in Eq.~(\ref{eq:LintVecTensMassEigs}) results from
integrating out a light mediator, processes in which the mediator is produced on
shell may provide stronger constraints than the model-independent interactions listed
above.  For this reason, we emphasize that the collider bounds presented 
in Figs.~\ref{fig:delmvsLambdaVECT} and~\ref{fig:delmvsLambdaTENS} should be interpreted
only heuristically and are only strictly valid when $\Lambda \gg \mathcal{O}(\tev)$.     
Similar caveats pertain to the co-annihilation constraints on our scenario, which 
are only strictly valid when $\Lambda \gg \mathcal{O}(m)$.  

Second, we note that while the recoil-energy thresholds for the direct-detection 
experiments relevant for this analysis are currently $\mathcal{O}({\rm keV})$,
future experiments could potentially achieve a comparable sensitivity 
with ${\cal O}({\rm eV})$ thresholds.  Such experiments would potentially be
able to discriminate between up-scattering, down-scattering, and elastic scattering
on the basis of the recoil-energy spectrum for mass splittings 
down to $\Delta m \sim \mathcal{O}({\rm eV})$.  For such small values of 
$\Delta m$, the lifetime of $\chi_2$ can be comparable to the age of the 
universe in the case of an antisymmetric-tensor interaction, implying that 
down-scattering might also be distinguishable at direct-detection experiments in
this case. 

Finally, in this analysis, we have only considered tree-level interactions between the 
dark and visible sectors of the form specified in Eq.~(\ref{eq:LintVecTensMassEigs}).  
However, we note that these tree-level operators also generically give rise to 
additional interactions at the loop level.  For example, operators arise at one loop 
which contribute to elastic scattering at direct-detection experiments.  While such process 
are expected to be suppressed, they can nevertheless be important in regions of parameter 
space within which other probes of the dark sector are insensitive.


\bigskip
\begin{acknowledgments}


We are happy to thank Danny Marfatia and Farinaldo Queiroz for discussions.
We also thank CETUP* (the Center for Theoretical Underground Physics and Related 
Areas) for hospitality during its 2015 Summer Program, where this work was initiated.
The research activities of KRD and DY are supported in part by 
the Department of Energy under 
Grant DE-FG02-13ER41976 (DE-SC0009913), 
while those of JK are supported in part by NSF CAREER grant PHY-1250573
and those of BT are supported
in part by NSF grant PHY-1720430.  
The research activities of KRD are also
supported in part by
the National Science Foundation through its employee IR/D program.
The opinions and conclusions expressed herein are 
those of the authors, and do not represent any funding agencies.

\end{acknowledgments}



\begin{thebibliography}{99}

\bibitem{DMReviews}
  For reviews, see, \eg:\\
  G.~Jungman, M.~Kamionkowski and K.~Griest,
  Phys.\ Rept.\  {\bf 267}, 195 (1996)
  [arXiv:hep-ph/9506380];\\
  J.~D.~Lewin and P.~F.~Smith,
  Astropart.\ Phys.\  {\bf 6}, 87 (1996);\\
  D.~Hooper,
  [arXiv:0901.4090 [hep-ph]]; \\
  N.~Weiner,
  ``Dark Matter Theory,'' video of lectures given at TASI 2009,
  {\tt http://physicslearning2.colorado.edu/}\\
  {\tt tasi/tasi\_2009/tasi\_2009.htm}; \\
  J.~L.~Feng,
  Ann.\ Rev.\ Astron.\ Astrophys.\  {\bf 48}, 495 (2010)
  [arXiv:1003.0904 [astro-ph.CO]];\\
  R.~W.~Schnee,
  arXiv:1101.5205 [astro-ph.CO];\\
  M.~Lisanti,
  arXiv:1603.03797 [hep-ph].
  
\bibitem{ComplementarityWhitePaper}
  S.~Arrenberg {\it et al.},
  arXiv:1310.8621 [hep-ph].

\bibitem{DecayComplementarity}
  K.~R.~Dienes, J.~Kumar, B.~Thomas and D.~Yaylali,
  Phys.\ Rev.\ Lett.\  {\bf 114}, no. 5, 051301 (2015)
  [arXiv:1406.4868 [hep-ph]].

\bibitem{InelasticDM} 
  D.~Tucker-Smith and N.~Weiner,
  Phys.\ Rev.\ D {\bf 64}, 043502 (2001)
  [hep-ph/0101138].

\bibitem{KoppSchwetzZupan} 
  J.~Kopp, T.~Schwetz and J.~Zupan,
  JCAP {\bf 1002}, 014 (2010)
  [arXiv:0912.4264 [hep-ph]].

\bibitem{Planck} 
  P.~A.~R.~Ade {\it et al.} [Planck Collaboration],
  Astron.\ Astrophys.\  {\bf 594}, A13 (2016)
  [arXiv:1502.01589 [astro-ph.CO]].

\bibitem{WessZumino} 
  J.~Wess and B.~Zumino,
  Phys.\ Lett.\  {\bf 37B}, 95 (1971).
  
\bibitem{WittenCurrentAlgebra} 
  E.~Witten,
  Nucl.\ Phys.\ B {\bf 223}, 422 (1983).

\bibitem{Kumarfatia}
  J.~Kumar and D.~Marfatia,
  Phys.\ Rev.\ D {\bf 88}, no. 1, 014035 (2013)
  [arXiv:1305.1611 [hep-ph]].

\bibitem{Das}
  A.~Das and B.~Dasgupta,
  Phys.\ Rev.\ Lett.\  {\bf 118}, no. 25, 251101 (2017)
  [arXiv:1611.04606 [hep-ph]].

\bibitem{BhattacharyaLattice}
  T.~Bhattacharya {\it et al.} [PNDME Collaboration],
  Phys.\ Rev.\ D {\bf 92}, no. 9, 094511 (2015)
  [arXiv:1506.06411 [hep-lat]].

\bibitem{HillSolon}
  R.~J.~Hill and M.~P.~Solon,
  Phys.\ Rev.\ D {\bf 91}, 043505 (2015)
  [arXiv:1409.8290 [hep-ph]].

\bibitem{SavageFreese}
  C.~Savage, G.~Gelmini, P.~Gondolo and K.~Freese,
  JCAP {\bf 0904}, 010 (2009)
  [arXiv:0808.3607 [astro-ph]].

\bibitem{MadGraph}
  J.~Alwall, M.~Herquet, F.~Maltoni, O.~Mattelaer and T.~Stelzer,
  JHEP {\bf 1106} (2011) 128
  [arXiv:1106.0522 [hep-ph]].

\bibitem{FeynRules1} 
  N.~D.~Christensen and C.~Duhr,
  Comput.\ Phys.\ Commun.\  {\bf 180}, 1614 (2009)
  [arXiv:0806.4194 [hep-ph]].

\bibitem{FeynRules2}
  A.~Alloul, N.~D.~Christensen, C.~Degrande, C.~Duhr and B.~Fuks,
  Comput.\ Phys.\ Commun.\  {\bf 185}, 2250 (2014)
  [arXiv:1310.1921 [hep-ph]].

\bibitem{Pythia}
  T.~Sjostrand, S.~Mrenna and P.~Z.~Skands,
  JHEP {\bf 0605} (2006) 026
  [hep-ph/0603175].

\bibitem{Delphes} 
  S.~Ovyn, X.~Rouby and V.~Lemaitre,
  arXiv:0903.2225 [hep-ph].

\bibitem{BoyarskyMWHalo}
  A.~Boyarsky, J.~Nevalainen and O.~Ruchayskiy,
  Astron.\ Astrophys.\  {\bf 471}, 51 (2007)
  [astro-ph/0610961].

\bibitem{BoyarskyXMMNewton}
  A.~Boyarsky, D.~Iakubovskyi, O.~Ruchayskiy and V.~Savchenko,
  Mon.\ Not.\ Roy.\ Astron.\ Soc.\  {\bf 387}, 1361 (2008)
  [arXiv:0709.2301 [astro-ph]].

\bibitem{MalyshevDwarfs}
  D.~Malyshev, A.~Neronov and D.~Eckert,
  Phys.\ Rev.\ D {\bf 90}, 103506 (2014)
  [arXiv:1408.3531 [astro-ph.HE]].

\bibitem{RuchayskiyXMMNewton}
  O.~Ruchayskiy {\it et al.},
  arXiv:1512.07217 [astro-ph.HE].

\bibitem{BoyarskyDecayBoundCluster}
  A.~Boyarsky, O.~Ruchayskiy and M.~Markevitch,
  Astrophys.\ J.\  {\bf 673}, 752 (2008)
  [astro-ph/0611168].

\bibitem{BoyarskyRuchayskiy}
  A.~Boyarsky and O.~Ruchayskiy,
  arXiv:0811.2385 [astro-ph].

\bibitem{QueirozOverview}
  F.~S.~Queiroz,
  arXiv:1605.08788 [hep-ph].

\bibitem{SterileNeutrinoBonds}
  M.~Drewes {\it et al.},
  [arXiv:1602.04816 [hep-ph]].

\bibitem{Hitomi} 
  F.~A.~Aharonian {\it et al.} [Hitomi Collaboration],
  Astrophys.\ J.\  {\bf 837}, no. 1, L15 (2017)
  [arXiv:1607.07420 [astro-ph.HE]].

\bibitem{SlatyerReionization}
  T.~R.~Slatyer and C.~L.~Wu,
  Phys.\ Rev.\ D {\bf 95}, no. 2, 023010 (2017)
  [arXiv:1610.06933 [astro-ph.CO]].

\bibitem{FermiLATDwarfSearch}
  M.~Ackermann {\it et al.} [Fermi-LAT Collaboration],
  Phys.\ Rev.\ Lett.\  {\bf 115}, no. 23, 231301 (2015)
  [arXiv:1503.02641 [astro-ph.HE]].

\bibitem{FermiBoundsWithMediators} 
  L.~M.~Carpenter, R.~Colburn, J.~Goodman and T.~Linden,
  Phys.\ Rev.\ D {\bf 94}, no. 5, 055027 (2016)
  [arXiv:1606.04138 [hep-ph]].

\bibitem{XENON1T} 
  E.~Aprile {\it et al.} [XENON Collaboration],
  arXiv:1705.06655 [astro-ph.CO].

\bibitem{LUXSI}
  D.~S.~Akerib {\it et al.} [LUX Collaboration],
  Phys.\ Rev.\ Lett.\  {\bf 118} (2017) no.2,  021303
  [arXiv:1608.07648 [astro-ph.CO]].

\bibitem{Helm} 
  R.~H.~Helm,
  Phys.\ Rev.\  {\bf 104}, 1466 (1956).
  
\bibitem{Engel} 
  J.~Engel,
  Phys.\ Lett.\ B {\bf 264}, 114 (1991).

\bibitem{PANDAX}
  X.~Cui {\it et al.} [PandaX-II Collaboration],
  arXiv:1708.06917 [astro-ph.CO].

\bibitem{LUXSD}
  D.~S.~Akerib {\it et al.} [LUX Collaboration],
  Phys.\ Rev.\ Lett.\  {\bf 118}, no. 25, 251302 (2017)
  [arXiv:1705.03380 [astro-ph.CO]].

\bibitem{PICO60} 
  C.~Amole {\it et al.} [PICO Collaboration],
  Phys.\ Rev.\ Lett.\  {\bf 118}, no. 25, 251301 (2017)
  [arXiv:1702.07666 [astro-ph.CO]].

\bibitem{KlosFormFactors}
  P.~Klos, J.~Menéndez, D.~Gazit and A.~Schwenk,
  Phys.\ Rev.\ D {\bf 88}, no. 8, 083516 (2013)
  Erratum: [Phys.\ Rev.\ D {\bf 89}, no. 2, 029901 (2014)]
  [arXiv:1304.7684 [nucl-th]].

\bibitem{PICO60LEfficiency} 
  C.~Amole {\it et al.} [PICO Collaboration],
  Phys.\ Rev.\ D {\bf 93}, no. 5, 052014 (2016)
  [arXiv:1510.07754 [hep-ex]].

\bibitem{LZTDR} 
  B.~J.~Mount {\it et al.},
  arXiv:1703.09144 [physics.ins-det].

\bibitem{PICO500}
  Scott Fallows, talk given at TAUPP 2017,
  {\tt https://indico.cern.ch/event/606690/}\\{\tt contributions/2623446/}.

\bibitem{CMSMonoJetOrHadWZNew} 
  CMS Collaboration [CMS Collaboration],
  CMS-PAS-EXO-16-048.

\bibitem{ATLASMonojet13TeVNeW} 
  The ATLAS collaboration [ATLAS Collaboration],
  ATLAS-CONF-2017-060.

\bibitem{ATLASMonoWZHadDec13TeV} 
  M.~Aaboud {\it et al.} [ATLAS Collaboration],
  Phys.\ Lett.\ B {\bf 763}, 251 (2016)
  [arXiv:1608.02372 [hep-ex]].

\bibitem{ATLASMonoPhoton13TeV}
  M.~Aaboud {\it et al.} [ATLAS Collaboration],
  Eur.\ Phys.\ J.\ C {\bf 77}, no. 6, 393 (2017)
  [arXiv:1704.03848 [hep-ex]].

\bibitem{CMSMonoPhoton13TeV}
  A.~M.~Sirunyan {\it et al.} [CMS Collaboration],
  arXiv:1706.03794 [hep-ex].

\bibitem{CMSMonoZLeptons13TeV}
  A.~M.~Sirunyan {\it et al.} [CMS Collaboration],
  JHEP {\bf 1703}, 061 (2017)
  [arXiv:1701.02042 [hep-ex]].

\bibitem{ATLASMonoHiggsbb}
  M.~Aaboud {\it et al.} [ATLAS Collaboration],
  Phys.\ Lett.\ B {\bf 765}, 11 (2017)
  [arXiv:1609.04572 [hep-ex]].

\bibitem{CMSMonoHiggs}
  A.~M.~Sirunyan {\it et al.} [CMS Collaboration],
  arXiv:1703.05236 [hep-ex].

\bibitem{ATLASMonoHiggsGammaGamma}
  M.~Aaboud {\it et al.} [ATLAS Collaboration],
  arXiv:1706.03948 [hep-ex].

\bibitem{BaiTait}
  Y.~Bai and T.~M.~P.~Tait,
  Phys.\ Lett.\ B {\bf 710}, 335 (2012)
  [arXiv:1109.4144 [hep-ph]].

\bibitem{Yuhsin}
  R.~Primulando, E.~Salvioni and Y.~Tsai,
  JHEP {\bf 1507}, 031 (2015)
  [arXiv:1503.04204 [hep-ph]].



\bibitem{Bainew} 
  Y.~Bai and T.~M.~P.~Tait,
  Phys.\ Lett.\ B {\bf 710}, 335 (2012)
  [arXiv:1109.4144 [hep-ph]].

\bibitem{Weinernew} 
  N.~Weiner and I.~Yavin,
  Phys.\ Rev.\ D {\bf 86}, 075021 (2012)
  [arXiv:1206.2910 [hep-ph]].

\bibitem{Izagnew} 
  E.~Izaguirre, G.~Krnjaic and B.~Shuve,
  Phys.\ Rev.\ D {\bf 93}, no. 6, 063523 (2016)
  [arXiv:1508.03050 [hep-ph]].

\bibitem{Codex} 
  M.~J.~Baker {\it et al.},
  JHEP {\bf 1512}, 120 (2015)
  [arXiv:1510.03434 [hep-ph]].

\bibitem{Codex2}
  M.~Buschmann, S.~El Hedri, A.~Kaminska, J.~Liu, M.~de Vries, X.~P.~Wang, F.~Yu and J.~Zurita,
  JHEP {\bf 1609}, 033 (2016)
  [arXiv:1605.08056 [hep-ph]].

\bibitem{Codex3} 
  S.~El Hedri, A.~Kaminska, M.~de Vries and J.~Zurita,
  JHEP {\bf 1704}, 118 (2017)
  [arXiv:1703.00452 [hep-ph]].


\bibitem{Buchnew} 
  O.~Buchmueller, A.~De Roeck, K.~Hahn, M.~McCullough, P.~Schwaller, K.~Sung and T.~T.~Yu,
  JHEP {\bf 1709}, 076 (2017)
  [arXiv:1704.06515 [hep-ph]].

\bibitem{Morrisseynew}
  D.~E.~Morrissey and A.~P.~Spray,
  JHEP {\bf 1406}, 083 (2014)
  [arXiv:1402.4817 [hep-ph]].

\bibitem{Izagnew2}
  E.~Izaguirre, Y.~Kahn, G.~Krnjaic and M.~Moschella,
  Phys.\ Rev.\ D {\bf 96}, no. 5, 055007 (2017)
  [arXiv:1703.06881 [hep-ph]].




\bibitem{DDM1}
  K.~R.~Dienes and B.~Thomas,
  Phys.\ Rev.\ D {\bf 85}, 083523 (2012)
  [arXiv:1106.4546 [hep-ph]].

\bibitem{DDM2}
  K.~R.~Dienes and B.~Thomas,
  Phys.\ Rev.\ D {\bf 85}, 083524 (2012)
  [arXiv:1107.0721 [hep-ph]].

\bibitem{DDM3}
  K.~R.~Dienes and B.~Thomas,
  Phys.\ Rev.\ D {\bf 86}, 055013 (2012)
  [arXiv:1203.1923 [hep-ph]].



\end{thebibliography}
\end{document}